\documentclass{iopjournal}

 \usepackage{tabularx}
 \usepackage{graphicx}
\usepackage{dcolumn}
\usepackage{array}
\usepackage{amsmath}
\usepackage{amssymb}

\newcolumntype{L}{>{\raggedright\arraybackslash}X}
\newcolumntype{Y}{>{\centering\arraybackslash}X}

\begin{document}

\articletype{Paper} 

\title{Models for photonic quantum networking with multi-layer cavities}

\author{Astghik Saharyan$^{1,*}$\orcid{0000-0002-1580-4103}, Juan-Rafael Álvarez$^2$\orcid{0000-0003-3411-8237}, Axel Kuhn$^3$\orcid{0000-0002-5101-8732} and St\'ephane Gu\'erin$^{4}$\orcid{0000-0002-9826-9598}}

\affil{$^1$Institut de Physique Nucléaire, Atomique et de Spectroscopie, CESAM, University of Liège, B-4000 Liège, Belgium}

\affil{$^2$LTCI, T\'el\'ecom Paris, Institut Polytechnique de Paris, 19 Place Marguerite Perey, 91120, Palaiseau, France}

\affil{$^3$University of Oxford, Clarendon Laboratory, Parks Road, Oxford, OX1 3PU, UK}

\affil{$^4$Laboratoire Interdisciplinaire Carnot de Bourgogne, CNRS UMR 6303, Universit\'e de Bourgogne,
BP 47870, 21078 Dijon, France}

\affil{$^*$Author to whom any correspondence should be addressed.}

\email{asaharyan@uliege.be}


\begin{abstract}
Effective models to describe the dynamics of an open cavity have been extensively discussed in the literature. In many of these models the cavity leakage to the outside is treated as a loss introduced phenomenologically. In contrast to these, we focus here on characterizing the outgoing photon using a novel approach where the outside is treated as part of the system. In such a global system, in order to separately characterize the photon inside and outside cavity, we demonstrate a first-principle derivation of a coherent cavity-reservoir coupling function for cavities with mirrors consisting of a stack of dielectric layers. In particular, we show that due to the effects induced by the multilayer nature of the cavity mirror, even in the standardly defined high-finesse cavity regime, the cavity-reservoir system description might differ from the one where the structure of the mirror is neglected. Based on this, we define a generalized cavity response function and a cavity-reservoir coupling function, which account for the longitudinal geometric structure of the cavity mirror. This allows us to define a resonance-dependent effective reflectivity for the cavity with a multilayer mirror as if it was sitting in a mode-dependent effective mirror plane. We estimate the error of such a definition by considering cavities of different lengths and mirror structures. Finally, we apply this model to characterize the dynamics of a single photon produced in such a cavity and propagating outside. 
\end{abstract}

\section{Introduction }
The realization of {distributed quantum processing, with stationary nodes coupled by flying qubits} relies on controllable interactions between light and matter at the few-photon level, allowing the teleportation of quantum states between distant nodes forming a quantum network \cite{nielsen00, Kimble2008,cQED2,ciracQuantumOpticsWhat2017a,Cacciapuoti1}. Systems such as atoms in cavity quantum electrodynamics (cavity QED)~\cite{Ritter2012,Gorshkov,McKeever,cQED4}, atomic ensembles~\cite{Jaksch2000,Saffman2010}, trapped ions~\cite{Keller2004,Stick2006} or quantum dots~\cite{Ding,Yao,Somaschi2016} are prime examples of such quantum nodes. In particular, cavity QED systems are one of the leading candidates for robust, controllable single and multiple photon sources, and they are featured in many proposals associated with distributed quantum information processing and communication~\cite{moehringEntanglementSingleatomQuantum2007,barrettMultimodeInterferometryEntangling2019a,cQED1,Wilk,Boozer,Dilley}. 
 
 A controllable, deterministic and reversible interaction between light and matter within the cavity can be achieved in the strong coupling regime, which implies the use of {high-finesse cavities formed by bespoke dielectric mirrors} to enhance {the interaction between the emitter and the electric field within a cavity} such that the coherent Rabi frequency of the atom-field interaction is faster than the spontaneous emission rate of the atom or the decay rate of the field in the cavity~\cite{Law}. {This is expressed by the system cooperativity, given by $2C=g^2/\kappa \gamma$, which in the strong coupling regime satisfies $2C \gg 1$. Here, $g$ describes the atom-cavity coupling, $\kappa$ characterizes the cavity field decay, and $\gamma$ corresponds to the atomic spontaneous emission.} Conversely, there is the weak coupling regime, in which there are significant losses present in the system. {These include the loss of coherence between the state of the emitter and cavity due to} the escape of the produced photon through the cavity mirrors. The latter, however, can be advantageous when considering quantum information extraction and transfer. While often considered a loss, the escape of photons through the cavity mirrors can be advantageous for quantum information processing, as it enables the extraction and transfer of quantum information. By operating in an intermediate coupling regime, the produced photon can be coupled to the output and transferred between distant quantum nodes~\cite{An_PhysRevA.79.032303,Barak1152261}. Such a propagating photon serves as a carrier of quantum information in a quantum network, linking spatially separated nodes. Thus, it is essential to develop effective models to study the properties and dynamics of the propagating photon outside the cavity, without treating it as a loss, which is the primary focus of the current manuscript.

In theoretical models, used to describe the dynamics of cavity QED systems, the optical and microwave cavities are usually modelled as isolated systems, the mirrors separating the field inside the cavity from the environment outside~\cite{Jaynes1963,Eberly1980}. Even when the cavity is not treated as perfect the interaction of the cavity with the environment is modelled as a weak damping of the cavity modes~\cite{BARNETT1988364}. One of the {methods} used to describe the dynamics of such systems is called input - output formalism~\cite{Dilley,gardiner00}. In this formalism, all of the properties of the field exiting the cavity (the output) can be determined based upon knowledge of the input field and the dynamics of the emitter-cavity system alone. These types of formulations are usually derived under the assumption that the environment spectrum is flat, i.e., the factor {$\kappa$ that describes} the damping of the cavity modes is a constant that is introduced phenomenologically~\cite{gardiner00,Breuer}. {In general, $\kappa$ depends on the frequency, mirror reflectivity, cavity length, transverse mode structure, and polarization. Variations in $\kappa$ for the cavity modes under consideration are largely disregarded in any phenomenological approach. Whenever a phenomenological approach is taken, the field modes inside the cavity are constraint to cavity eigenmodes. Often a single eigenmode is relevant, and its time-dependent probability amplitude is proportional to the field amplitude of the emitted photon. 
The a-priori restriction to cavity eigenfrequencies is highly problematic if the frequency of the light emitted by the atom is substantially different. In that case, any holistic model, using a continuum of modes spanning cavity and outside world, is better suited to describe the physical process and to eventually calculate the photon as temporal shape and spectrum.}

In the literature, these issues have been discussed, and the construction of models {justifying phenomenological ones or divergent from them, derived from first principles, has been proposed. The true-mode (or mode-of-the-universe) representation is the most general form where the field is quantized from the Maxwell equations and the (generally continuous) eigenmodes of the total system comprising the cavity and the outside 
~\cite{Dutra2000,dutra2005cavity,pseudomodes}. As demonstrated in Refs.~\cite{Dutra2000,dutra2005cavity}, one can then separate these true modes into inside and outside cavity modes and derive an approximate coherent cavity-environment coupling function for cavities with sufficiently low transmissivity.  
The procedure has been formally generalized in \cite{Viviescas,Lentrodt}, where a microscopic derivation of the field quantization in open cavities with mirrors of arbitrary shape is established and the system-bath Hamiltonian is derived.
This is closely related to the pseudomode method \cite{pseudomodes} which aims at treating a non-Markovian dynamics with a master equation of Lindblad form \cite{Tamascelli,Medina} or not \cite{Pleasance} by mapping the system onto an auxiliary configuration comprising small numbers of discret modes each coupled to an independent Markovian reservoir. These generalizations are particularly useful for nanostructures involving plasmonics (see e.g. \cite{Medina}), whose modelling has however raised controversies \cite{Semin}.}

{In this paper, we aim at analyzing single-photons produced in and leaking out from a high-finesse multi-layer cavity coupled to a driven atom in a configuration close to the experiment. We therefore deem appropriate to adopt an inside-outside representation with quasi-modes, as a specific configuration of the general procedure \cite{Viviescas,Lentrodt}, rather than using pseudomodes with a master equation (following the nomenclature of Ref. \cite{pseudomodes}) or quasinormal modes \cite{Franke1,Franke2} whose quantization remains challenging due to their lossy nature.  }

True-mode, inside-outside, and pseudomode approaches allow to test the accuracy of the phenomenological models. In Table~\ref{tabel:formalisms}, we summarize their corresponding validity limits, as discussed in the literature, including the results of this paper~\cite{Saharyan5.033056,dutra2005cavity,pseudomodes,vogel2006quantum}. As discussed in Ref.~\cite{Saharyan5.033056}, analyzing the dynamics using such a coherent coupling function (referred to as the inside-outside formalism) inherently provides complete characteristics of the produced photon both inside and outside the cavity.  Having such a complete description of the entire network, including the photon is, in principle, sufficient for characterizing the quantum network between distant quantum nodes. Furthermore, employing this formalism allows for the definition of parameters with clear physical meanings and enables the derivation of explicit validity limits for the approximate model~\cite{Saharyan5.033056}. Additionally, one can derive the pseudomode representation from both the inside-outside and true-mode representations. This representation is analogous to phenomenological models, wherein the escape of the photon from the cavity is treated as a constant loss. Unlike phenomenological models, in the pseudomode representation, this loss rate and the validity limit of this representation can be explicitly estimated~\cite{Saharyan5.033056,pseudomodes}.
\begin{table}[h!]
\begin{center}
\begin{tabularx}{\columnwidth}
{|>{\hsize=0.9\hsize}L|
 >{\hsize=1\hsize}Y|
 >{\hsize=1.1\hsize}Y|}
\hline 
 Formalism &  \multicolumn{2}{c}{Validity limit} \vline \tabularnewline
\hline 
{} &   Single layer &  Multilayer \tabularnewline
\hline 
\hline 
 True-mode  \cite{Dutra2000,dutra2005cavity,vogel2006quantum}  & {$\infty$} &  {$\infty $}\tabularnewline
\hline 
 Inside-outside \cite{Dutra2000,dutra2005cavity,Saharyan5.033056}  &  {${\Gamma_{c}\ell_{c}}/{c}\ll1$}  & {${{\kappa}_{m}\ell_{\text{eff}}}/{c}\ll1,\; \zeta_m\ll1$} \tabularnewline
\hline 
 Pseudomode \cite{pseudomodes,Saharyan5.033056,multilayer} & ${\Gamma_{c}(x_{A}+\ell_{c})}/{c}\ll1$  & ${{\kappa}_{m}(x_{A}+\ell_{c})}/{c}\hspace{-0.1cm}\ll1$ \tabularnewline
\hline 
\end{tabularx}
\end{center}
\caption{Summary of different formalisms used to describe the dynamics of an atom trapped in a leaky optical cavity and their validity limits, where the limit for the inside-outside representation for the multilayer case is derived in this work. $\Gamma_{c}$ and ${\kappa}_{m}$ are the cavity linewidths for cavity mirror made of single and multilayer dielectrics respectively, and $\ell_{c}$ and $\ell_{\text{eff}}$ are the corresponding cavity lengths. When considering more realistic cavities with multilayer dielectric mirrors, as we show in this work, these conditions feature the multilayer cavity parameters.}
\label{tabel:formalisms}
\end{table}

{In the derivations of the approximate models discussed above, the structure of the cavity mirrors is commonly disregarded, treating the mirrors as a dielectric with a negligible thickness. As can be seen from the validity limits of these models (Table~\ref{tabel:formalisms}), for them to work, the cavity must be of high finesse, which in reality can not be ensured with a single {transition between} dielectric layers.} If we consider Fabry-P\'erot type cavities, in practice the high finesse is achieved via increasing the number of dielectric layers forming the cavity mirrors. As shown in Ref.~\cite{multilayer}, the dielectric structure of the cavity mirrors may alter the characteristic parameters of the cavity, such as cavity free spectral range, mode volume and the effective length determining the cavity resonances. 

Taking into account these studies, here we extend the inside--outside representation established in Ref.~\cite{Saharyan5.033056} to cavities with realistic multilayer dielectric mirrors, building on the multilayer cavity response developed in Ref.~\cite{multilayer}. Such an extension allows us to estimate the effects of the {longitudinal} geometric structure of the cavity on the coherent cavity-reservoir coupling function and on the dynamics, focusing particularly on the dynamics of the outgoing photon for quantum networking purposes. To this aim we introduce a generalized cavity response function and an effective reflectivity that allow to describe the behavior of the multilayer mirror in terms of a single-layered one, whose characteristic parameters account for the actual multilayer nature of the structure. We estimate the error of such a reformulated description with respect to the number of dielectric layers and the geometric length of the cavity. We then follow the derivation scheme described in Ref.~\cite{dutra2005cavity}, where the cavity-reservoir coupling is derived by comparing the {true} modes of a closed system consisting of an open cavity and its surroundings to the {inside-outside} modes of a system where a perfect cavity is coupled to the semi-infinite reservoir of continuous modes. This allows us to derive a generalized cavity-reservoir coupling function that accounts for the effects caused by the actual structure of the mirror. Finally, we apply this {inside-outside} model to study the dynamics of an atom trapped in such a multilayer cavity and study the properties of a photon produced from such a system. In particular, we demonstrate that apart from having the time profile of the produced photon, with this formulation, one can {directly} obtain the spectral shape of the {outgoing} photon as well as its spatial distribution.

The paper is organized as follows, in Section~\ref{sec:truemode}, we analyze the response function of the cavity with multilayer mirror and the limit at which the modes of such a structure can be represented as a sum of individual modes of a single-layered structure. We then write the electromagnetic field in the true-mode representation, i.e., for a closed system consisting of the multilayer cavity and its surroundings. In Section~\ref{sec:inout}, we analyze the model in the inside-outside representation, featuring a perfect cavity coupled to a semi-infinite reservoir. Finally, by imposing equivalence between the two representations we obtain the cavity-reservoir coupling in Section~\ref{sec:coupling}. In Section~\ref{sec:verification}, we study the dynamics of an atom trapped in such a multilayer cavity in an intermediate coupling regime, which allows the produced photon to escape the cavity.

\section{True-mode representation\label{sec:truemode}}

We start by analyzing a global closed system consisting of a non-perfect cavity with a partially transparent mirror and its surroundings. The quantization based on such a closed system is often referred to as universal quantization or true-mode representation~\cite{Knoll, dutra2005cavity, pseudomodes, murray1978laser, BARNETT1988364,Saharyan5.033056}. In this representation, one can obtain the {continuum of} normal modes by solving the classical boundary-value problem for a given structure of the cavity. Thus, one can quantize the model according to these modes that extend over inside as well as outside the cavity~\cite{multilayer}. Here, we consider the same structure of the cavity as the one presented in \cite{multilayer}, i.e., a one-dimensional cavity made of a perfect mirror on the left and a partially transparent flat mirror on the right, which consists of a stack of dielectric layer pairs (Fig.~\ref{fig:scheme}(a)). Since, in practice, the considered mirror structures are either flat or have large curvature radii, the field is described by its longitudinal propagation, while the transverse mode structure is incorporated through the normalization factor $\mathcal{A}$ defined below (paraxial approximation) In particular, for the fundamental paraxial TEM$_{00}$ mode of a stable cavity with appropriately curved mirrors, the wavefront curvature approximately matches that of the mirror surfaces, resulting in predominantly normal incidence. The angular and polarization dependence of the multilayer reflectivity can therefore be neglected, provided that the resulting corrections remain small compared with the relevant cavity linewidth. Consequently, this 1D model adequately captures the properties of realistic cavities used in the majority of the current experiments. The perfect mirror and the partially transparent mirror are placed at $x = -\ell_{c}$ and $x = 0$, respectively, forming a cavity of geometric length $\ell_{c}$. The mirror is designed to reflect best at $\lambda_{0}$ with the corresponding resonance frequency $\omega_{c}=2\pi c/\lambda_{0}$. As shown in Ref.~\cite{multilayer}, the modes for such a system can be described by the following functions:
\begin{subequations}
\label{eq:modes_in_out}
\begin{align}
\label{eq:mode_in}
\Phi_{\omega,\text{ins}}(x)&=\frac{2i}{\sqrt{2\pi c \mathcal{A}}}e^{i\frac{\omega}{c}\ell_{c}}\mathcal{T}_{\omega}\sin{\left[ \frac{\omega}{c}\left( x+\ell_{c}\right)\right]},\\
\label{eq:mode_out}
\Phi_{\omega, \text{outs}}(x)&=\frac{1}{\sqrt{2\pi c \mathcal{A}}}\left(e^{2i\frac{\omega}{c}\ell_{c}}\frac{\mathcal{T}_{\omega}}{\mathcal{T}^{\ast}_{\omega}}e^{i\frac{\omega}{c}x}-e^{-i\frac{\omega}{c}x} \right),
\end{align}
\end{subequations}
where $\Phi_{\omega,\text{ins}}(\omega)$ is the mode corresponding to the inside of the cavity, specified by the interval $-\ell_c<x<0$, and $\Phi_{\omega,\text{outs}}(x)$ is the mode relating to the outside of the cavity, corresponding to the interval $s < x< \infty$, where $s$ is the position where the multilayer stack ends. The inclusion of modes $\Phi_{\omega,\text{stack}}(\omega)$ within the stack is not relevant in this work; the explicit expression can be found in Ref.~\cite{multilayer}. Here, $\mathcal{A}$ is the transverse area of the mode, and $\mathcal{T}_{\omega}$ is the multilayer cavity response function that describes the intensity ratio of light inside and outside the cavity (see the explicit form in Appendix~\ref{app:equivalcne_single_layer}, Eq.~\eqref{eq:resp_actual}). As demonstrated in Appendix~\ref{app:equivalcne_single_layer}, the exact response obtained from the recursive boundary-matching formulation can equivalently be expressed in terms of the complex reflection and transmission coefficients of the complete multilayer stack,
\begin{align*}
\mathcal{T}_{\omega}=\frac{t_{\mathrm{multi}}(\omega)}{1+r_{\mathrm{multi}}(\omega)e^{2i\omega\ell_{\mathrm{ref}}/c}},
\end{align*}
where $\ell_{\mathrm{ref}}=\ell_c+d_{\mathrm{stack}}/2$, with $d_{\mathrm{stack}}$ denoting the physical thickness of the dielectric stack. This representation is equivalent to the conventional transfer-matrix description, while the recursive formulation additionally provides the spatial true modes required for field quantization. At the limit of high finesse cavity, insuring sufficiently narrow and isolated resonances, the response function can be written in the following form \cite{multilayer}:
\begin{subequations}
\label{eq:resp}
\begin{align}
\label{eq:multilayer_resp}
\mathcal{T}_{\omega}& \approx\sum_{m} \mathcal{T}_{m}(\omega), \\
\label{eq:multilayer_resp_ind}
 \mathcal{T}_{m}(\omega) &= \sqrt{\frac{c}{2L^{(m)}_{N}}} \frac{\sqrt{{\kappa}^{(m)}_{N}}}{\left( \omega - \omega^{(m)}_{N}\right)+i\frac{{\kappa}^{(m)}_{N}}{2}},
\end{align}
\end{subequations}
where ${\kappa}^{(m)}_{N}$ is the cavity linewidth {that depends on the number of dielectric layer pairs used}. The index $m$ spans through $2{p}-1+2N+ \lceil 2(\ell_{c}-{p}\frac{\lambda_{0}}{2}) \rceil$, with ${p}$ being the number of antinodes within the length $\ell_{c}$ and $N$ being the number of dielectric pairs forming the partially transparent mirror. The presence of three different length parameters $L^{(m)}_{N}, \, \ell^{(m)}_{\text{eff}}$ and $\ell_{c}$ is due to the fact that in this model the structure of the mirror is explicitly taken into account \cite{multilayer}, i.e., in standard models, these three parameters are the same and correspond to the mirror spacing, $\ell_c$. Here, however, the three length parameters have distinct physical meanings: $\ell_c$ is the geometric cavity length, $\ell_{\mathrm{eff}}^{(m)}$ is the resonance-defined effective length satisfying $\omega_N^{(m)}=m\pi c/\ell_{\mathrm{eff}}^{(m)}$, and $L_N^{(m)}$ determines the normalization of the corresponding Lorentzian contribution. As shown in Appendix~\ref{app:equivalcne_single_layer}, within the isolated-resonance approximation, the latter is related to the frequency derivative of the physical multilayer reflection phase through
\begin{align*}
L_N^{(m)}\simeq \ell_{\mathrm{ref}}+\frac{c}{2}\left.\frac{d\varphi_{\mathrm{multi}}(\omega)}{d\omega}\right|_{\omega_N^{(m)}}.
\end{align*}
Thus, the Lorentzian parameters of eq.~\eqref{eq:multilayer_resp_ind} account not only for the magnitude of the multilayer reflectivity but also, to first order, for its frequency-dependent phase. Higher-order phase dispersion is not included in this local Lorentzian approximation.

\begin{figure}[!h]
\centering
\includegraphics[scale=0.6]{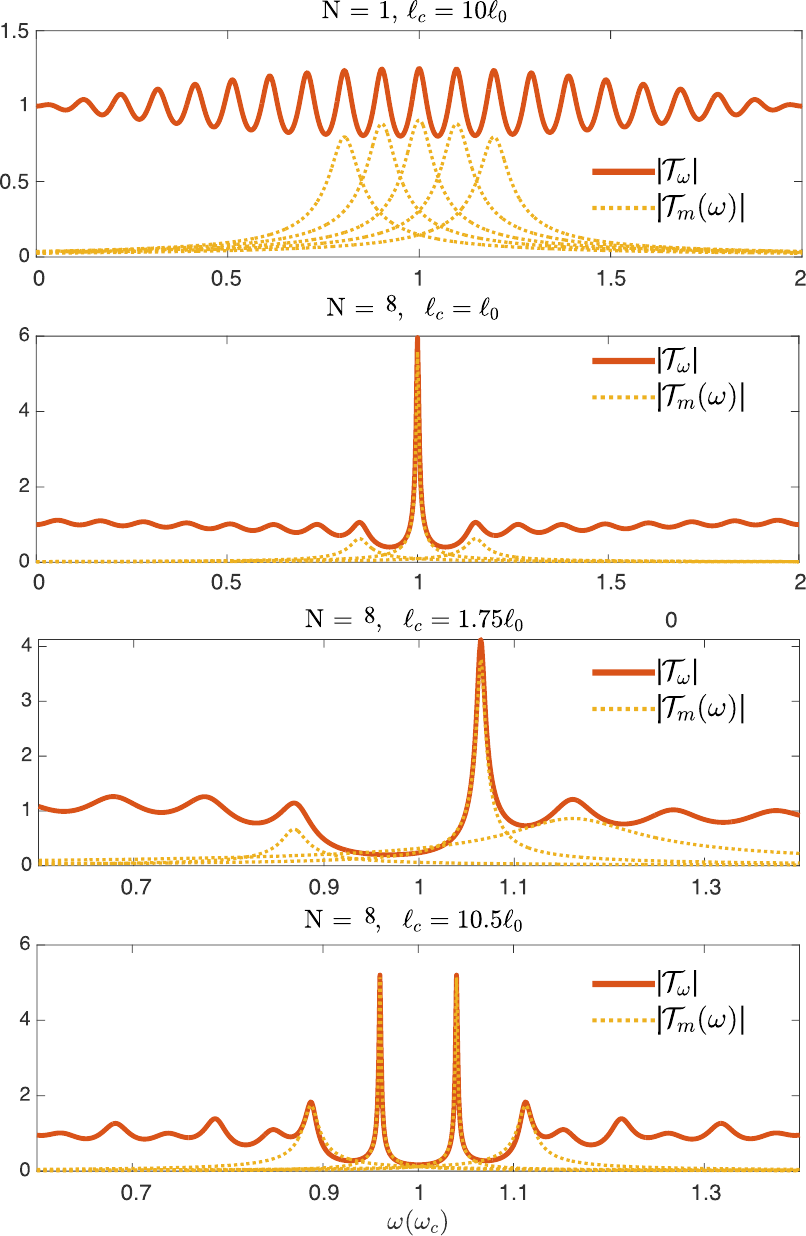}
\caption{Response function of a  cavity with multilayer dielectric mirror. Solid lines represent the actual response function $|\mathcal{T}_{\omega}|$, while dashed lines correspond to each term $|\mathcal{T}_{m}(\omega)|$ close to the resonance frequency $\omega_{c}$. $N$ is the number of dielectric pairs and $\ell_{c}$ is the mirror spacing, with $\ell_{0}=\lambda_{0}/2$.}
\label{fig:fig1}
\end{figure}

For the multilayer structure considered here, the parameters of the individual Lorentzian contributions are obtained numerically by imposing $|\mathcal{T}_{\omega}| \approx |\sum_m \mathcal{T}_{m}(\omega)|$, yielding the fitted parameters $\kappa_N^{(m)}$, $\omega_N^{(m)}$, and $L_N^{(m)}$. In the regime of narrow, well-isolated resonances, these parameters can be related to the approximate analytical expressions derived in Appendix~\ref{app:equivalcne_single_layer}. In Fig.~\ref{fig:fig1}, we illustrate the response function $\mathcal{T}_{\omega}$ [Eq.~\eqref{eq:resp_actual} of Appendix~\ref{app:equivalcne_single_layer}] and the individual terms $\mathcal{T}_{m}(\omega)$ for cavities with different mirror spacing and dielectric layers. As we can see from the figure, for certain scenarios, the overlap between the neighboring $\mathcal{T}_{m}(\omega)$ is significant, and given the representation of finesse as the ratio of cavity free spectral range to its linewidth, the significant overlap implies that the cavity is of low finesse. More generally, since the resonance frequencies and linewidths of a multilayer cavity are mode-dependent, the validity of the single-resonance approximation can be characterized by the parameter
\begin{align}
\zeta_m=\max_{j=m\pm1} \frac{\kappa_N^{(m)}+\kappa_N^{(j)}}{2|\omega_N^{(m)}-\omega_N^{(j)}|}.
\end{align}
The condition $\zeta_m\ll1$ provides a practical criterion for negligible spectral overlap between adjacent resonances and reduces to the usual high-finesse condition for approximately equally spaced resonances with comparable linewidths. In addition, an accurate Lorentzian description requires that the multilayer reflection magnitude vary slowly and its round-trip phase remain approximately linear across the selected resonance.

\begin{figure}[!h]
\centering
\includegraphics[width=0.6\linewidth]{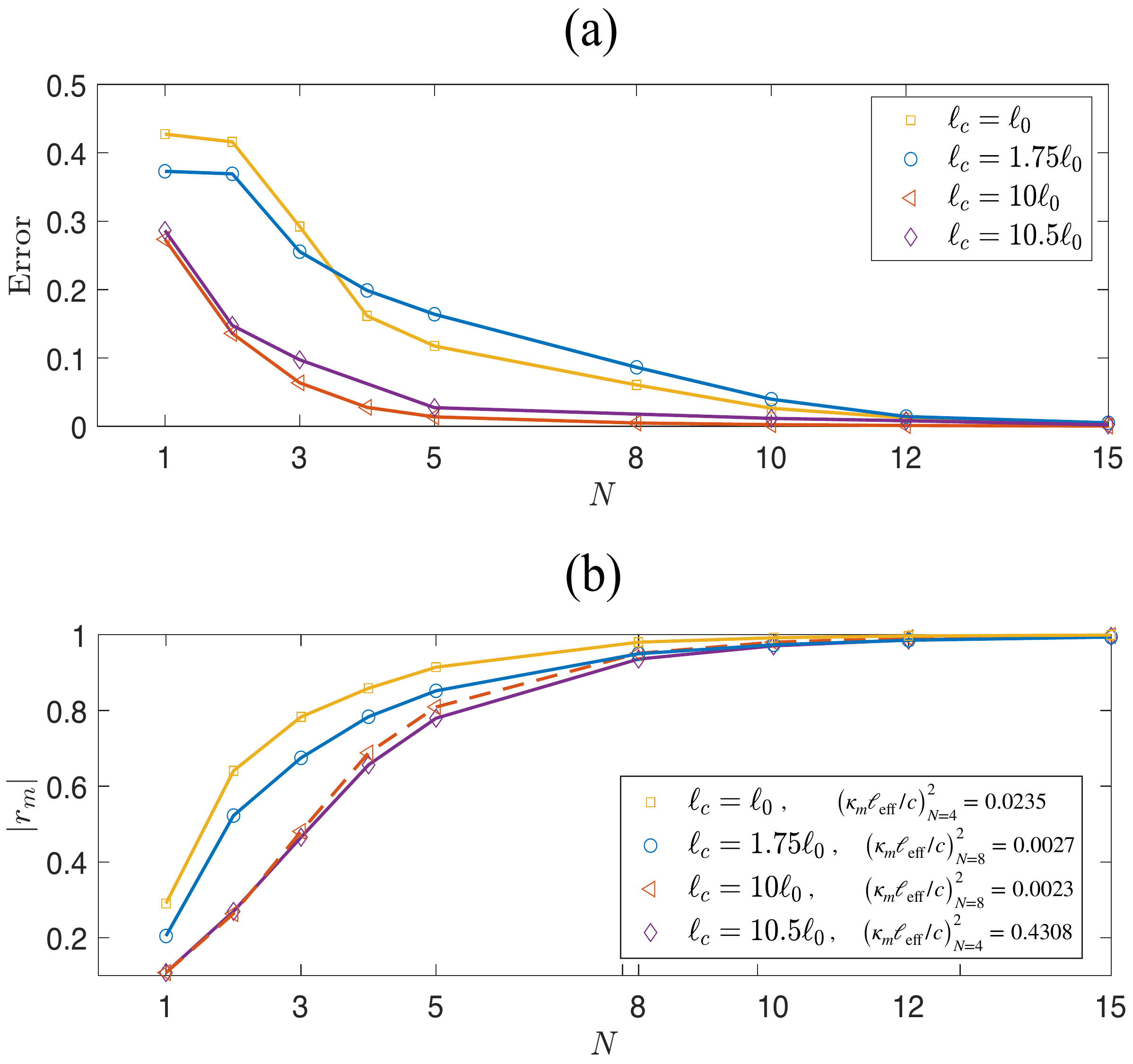}
\caption{(a): The error of approximating the full response function by a single term of the form of Eq.~\eqref{eq:multilayer_resp_ind} around the resonance peak $\omega_{c}$ for cavities with different lengths $\ell_{c}$ and with different numbers of dielectric layer pairs $N$. (b): Effective reflectivity of the mirror replacing the partially transparent multilayer mirror: $|r_{m}| =r_{m}(\omega^{(m)}_{N}) = e^{-{\kappa}^{(m)}_{N}\ell^{(m)}_{\rm{eff}}/c}$. }
\label{fig:fig2}
\end{figure}

For the situations where the overlap is negligible, around the resonance peak $\omega_{c}$ the full response function~\eqref{eq:multilayer_resp} can be reduced to a single term $\mathcal{T}_{m}(\omega)$: $\mathcal{T}_{\omega} \approx \mathcal{T}_{m}(\omega)$ (high-finesse). In particular, in Fig.~\ref{fig:fig2}(a), we show the error of such an approximation for cavities of different mirror spacing $\ell_{c}$ and dielectric structure. The error is estimated by calculating the term
 \begin{eqnarray}
  1 - \Big | \frac{\langle \mathcal{T}_{\omega},  \mathcal{T}_{m}(\omega) \rangle }{\langle  \mathcal{T}_{\omega},  \mathcal{T}_{\omega}  \rangle } \Big |=  1 - \Big | \frac{\int d\omega \, \mathcal{T}_{\omega}  \mathcal{T}^{\ast}_{m}(\omega)  }{\int d\omega  |\mathcal{T}_{\omega}|^{2} } \Big |,
  \end{eqnarray}
where the integration is done over the region $\omega^{(m)}_{N} - {\kappa}^{(m)}_{N}/2 < \omega < \omega^{(m)}_{N}+ {\kappa}^{(m)}_{N}/2$. As we can see, {when the number of layers is small, because there are significant overlaps between individual Lorentzians, the error of the approximation is relatively high. Additionally,} for the cases where the {length of the cavity is such that} $\ell_{c}/\ell_{0} \neq {p}$, the main resonance peak is not symmetric with respect to the neighboring peaks, and while it is well separated from one, it can be significantly overlapped with the other, increasing the error in the approximation.

Taking into account this analysis of the response function, in the following, we reformulate $\mathcal{T}_{m}(\omega)$ around each peak $\omega^{(m)}_{N}$  in terms of a single-layered cavity response function derived in Refs.~\cite{vogel2006quantum, dutra2005cavity}. This reformulation can be thought of as replacing a multilayer mirror with an effective single-layered one with negligible thickness, which is positioned such that it forms a cavity of mirror spacing $\ell^{(m)}_{\rm{eff}}$ and the resonance frequency $\omega^{(m)}_{N}$. The spectral transmission and reflection functions for this fictitious mirror are $t_{m}(\omega)$ and $r_{m}(\omega)$ respectively, such that (see the details in Appendix~\ref{app:equivalcne_single_layer}):
\begin{eqnarray}
\label{eq:resp_sing_for_mult}
\mathcal{T}_{m}(\omega) \approx T_{m,\omega}= \sqrt{\frac{\ell^{(m)}_{\text{eff}}}{L^{(m)}_{N}}}\frac{t_{m}(\omega)}{1+r_{m}(\omega)e^{2i\frac{\omega}{c}\ell^{(m)}_{\text{eff}}}},
\end{eqnarray}
with $|r_{m}(\omega^{(m)}_{N})| = e^{-{\kappa}^{(m)}_{N}{\ell^{(m)}_{\text{eff}}}/{c}}$, $r_{m}(\omega)=|r_{m}(\omega)|e^{i\varphi_{r}(\omega)}$ and $\varphi_{r}(\omega^{(m)}_{N})=\pi$. Both the effective mirror position and its resonant reflection coefficient are fixed by the selected resonance: the effective length $\ell_{\mathrm{eff}}^{(m)}$ is determined by the resonance frequency, the reflection magnitude by the linewidth, and the reflection phase is chosen to be $\pi$ to satisfy the effective-cavity resonance condition. As detailed in Appendix~\ref{app:equivalcne_single_layer}, the effective reflection magnitude can also be related to that of the physical multilayer mirror through $|r_m(\omega_N^{(m)})|\simeq |r_{\mathrm{multi}}(\omega_N^{(m)})|^{\ell_{\mathrm{eff}}^{(m)}/L_N^{(m)}}$. The effective mirror position is therefore not an independently adjustable parameter within this construction. The expression~\eqref{eq:resp_sing_for_mult} is different from the standard single-layered cavity response function by the factor $\sqrt{{\ell^{(m)}_{\text{eff}}}/{L^{(m)}_{N}}}$, which tends to one only for long {cavities that satisfy $\ell_c = {p}\ell_0$ (i.e., ${p}$ is a large integer number)} and are made of mirrors with large number of dielectric layers~\cite{multilayer}. Thus, Eq.~\eqref{eq:resp_sing_for_mult} can be considered as the generalized cavity response function, that incorporates the effects induced by the {longitudinal} structure of the cavity.
 Additionally, we can interpret the term $|r_{m}|:=|r_{m}(\omega^{(m)}_{N})|$ as the effective reflectivity of the multilayer cavity, and in Fig.~\ref{fig:fig2}(b), we show the dependence of this effective reflectivity on the number of dielectric layer pairs and the mirror spacing of the actual cavity. Comparing Fig.~\ref{fig:fig2}(b) with the error figure Fig.~\ref{fig:fig2}(a), we can estimate the value of the mirror reflectivity for which the complete spectral decomposition of modes can be replaced by individual ones. Finally, we emphasize that, although the above construction reproduces the local spectral intensity, it does not uniquely determine the full frequency-dependent complex coefficients $r_m(\omega)$ and $t_m(\omega)$ of the fictitious mirror, which should not be identified with those of the physical multilayer stack.

 \subsection{Electromagnetic field in the true-mode representation}

Using the modes in Eq.~\eqref{eq:modes_in_out}, we can write the quantized expressions of the electric and magnetic fields in true-mode representation~\cite{Federico_2022}. In particular, the fields inside and outside the cavity read as follows:
\begin{subequations}
\label{eq:real_model_electric}
\begin{align}
\nonumber
{E}_{\text{ins}}(x) =& -i\int_0^{\infty} d\omega \sqrt{\frac{\hbar \omega}{2\epsilon_0}}\left( \Phi_{\omega,\text{ins}}(x) {a}_{\omega} - \Phi_{\omega,\text{ins}}^{\ast}(x) {a}_{\omega}^{\dagger}\right)\\
=& \int_0^{\infty} d\omega \sqrt{\frac{\hbar\omega}{\pi c\mathcal{A}\epsilon_0}} \Big(\sin\left[{\textstyle\frac{\omega}{c}}(x+\ell_c)\right]e^{i\frac{\omega}{c}\ell_c}\mathcal{T}_{\omega}{a}_{\omega} +H.c.\Big),\\ \nonumber
{E}_{\text{outs}}(x) =& -i\int_0^{\infty} \hspace{-0.25cm} d\omega \sqrt{\frac{\hbar\omega}{2\epsilon_0}}\left( \Phi_{\omega,\text{outs}}(x) {a}_{\omega} - \Phi_{\omega,\text{outs}}^{\ast}(x) {a}_{\omega}^{\dagger}\right)\\ \label{eq:Eout}
=&-i\int_0^{\infty} d\omega \sqrt{\frac{\hbar \omega}{ 4\pi c \mathcal{A} \epsilon_{0}}}\Big( \Big [e^{2i\frac{\omega}{c}\ell_c}\dfrac{\mathcal{T}_{\omega}}{\mathcal{T}^*_{\omega}}e^{i\frac{\omega}{c}x}-e^{-i\frac{\omega}{c}x} \Big ] {a}_{\omega} - H.c. \Big),
\end{align}
\end{subequations}
\begin{subequations}
\label{eq:real_mode_magnetic}
\begin{align}
{B}_{\text{ins}}(x)=&\,\frac{i}{c}\int_0^{\infty} d\omega \sqrt{\frac{\hbar\omega}{\pi c\mathcal{A}\epsilon_0}} \Big( \cos\left[{\frac{\omega}{c}}(x+\ell_c)\right] e^{i\frac{\omega}{c}\ell_c}\mathcal{T}_{\omega}{a}_{\omega} - H.c. \Big),\\ \label{eq:Bout}
{B}_{\text{outs}}(x)=&\,\frac{i}{c}\int_0^{\infty} d\omega \sqrt{\frac{\hbar\omega}{4\pi c\mathcal{A}\epsilon_0}} \Big( \Big[e^{2i\frac{\omega}{c}\ell_c}\dfrac{\mathcal{T}_{\omega}}{\mathcal{T}^{\ast}_{\omega}}e^{i\frac{\omega}{c}x}+e^{-i\frac{\omega}{c}x} \Big ] {a}_{\omega} - H.c. \Big),
\end{align}
\end{subequations}
where ${a}_{\omega}$ is the annihilation operator of the true modes describing the whole cavity - environment closed system~\cite{Saharyan5.033056}. This operator satisfies the commutation relations
\begin{subequations}
\label{eq:commutator_global}
\begin{eqnarray}
\left[ {a}_{\omega}, {a}^{\dagger}_{\omega'}\right]&=&\delta(\omega - \omega'),\\
\left[ {a}_{\omega}, {a}_{\omega'}\right]&=&0.
\end{eqnarray}
\end{subequations}
Furthermore, the Hamiltonian for such a closed system in the true-mode representation writes 
\begin{align}
\label{eq:global_Hamiltonian}
{H}=\int_{0}^{\infty}d\omega ~ \hbar \omega  {a}^{\dagger}_{\omega}{a}_{\omega},
\end{align}
{where we {disregard} the term $\frac{1}{2}\int^{\infty}_{0} d\omega \hbar \omega$, with the argument that only differences in energy are physically measurable, thus the zero of the energy can be chosen at will in non-relativistic quantum physics. We highlight that retaining this term would result in an infinite quantity, and neglecting it is equivalent to applying a renormalization to eliminate such infinities.}

After writing the modes in the true-mode representation, in the following we separate these true modes into inside and outside ones.

\section{Inside - outside representation\label{sec:inout}}

In the previous section, we introduced the model in the true-mode representation, where the continuum of modes describes the full cavity-environment closed system. We now apply the inside–outside representation established in Ref.~\cite{Saharyan5.033056} to the multilayer cavity defined above. In this representation the inside modes correspond to the discrete modes of a perfect cavity and the outside modes are the continuous reservoir modes that are delimited by the perfect cavity. This separation allows one to study the dynamics of the cavity separated from its environment while still providing the characterization of the full closed system. Such a complete characterization then allows one to study the propagation of the produced photon outside the cavity, for information transfer purposes. The inside–outside representation was derived from first principles in Ref.~\cite{Saharyan5.033056}; here we determine how the presence of the finite-thickness multilayer mirror modifies the quantities entering that representation and its validity limit.  We note that this separation does not imply that the photon first exists entirely within the cavity before appearing outside.  Rather, it allows for a distinction in the distribution of the photon between the inside and outside of the cavity. In practice, photons produced by cavity QED systems have long temporal profile due to the timing control in the adiabatic passage. They are never completely in the cavity. Photon durations typically exceed the cavity decay time by an order of magnitude, and their spectral width is therefore also an order of magnitude narrower than the cavity linewidth.

In the derivation below, we consider a regime with relatively high effective reflectivity, so that we can approximate the full response function with a single term $\mathcal{T}_{m}(\omega)$, corresponding to the strongest resonance peak. Since we consider a single peak $m$, for readability, here after we drop the indexing $m$ for the terms $\ell^{(m)}_{\rm{eff}}:=\ell_{\rm{eff}}$ and $L^{(m)}_{N}: =L_{N}$, and redefine the resonance frequency $\omega_{m}:=\omega^{(m)}_{N}$ and the cavity linewidth ${\kappa}^{(m)}_{N}={\kappa}_{m}$. We note that if the longitudinal structure of the cavity were not taken into account, and it was assumed that the strongest peak corresponds to the resonance $\omega_m = m \pi c/\ell_c$, then in certain scenarios (e.g., the case $N=8$, $\ell_c = 1.75\ell_0$ in Fig.~\ref{fig:fig1}), $\mathcal{T}_{m}(\omega)$ would not align with the strongest resonance peak. Consequently, both the light-matter coupling inside the cavity and the below derived coupling of the photon to the outside would be incorrectly estimated.

\subsection{Electromagnetic field in the inside-outside representation}

The electric and magnetic fields in a perfect cavity of length $\ell_{\text{eff}}$, the left mirror of which is placed at $-\ell_{c}$ write as follows: 
\begin{subequations}
\label{eq:modes_inside}
\begin{align}
\mathbb{E}_{\text{ins}}(x)&=\sum_{m=1}^{\infty} \hspace{-0.1cm}\sqrt{\frac{\hbar\omega_m}{\ell_{\text{eff}}\mathcal{A}_{r}\epsilon_0}}\Big({a}_{m} + {a}^{\dagger}_{m}\Big)\sin{\Big [\frac{\omega_m}{c}(x+\ell_{c})\Big]},\\
\mathbb{B}_{\text{ins}}(x)&=\frac{i}{c}\hspace{-0.05cm}\sum_{m=1}^{\infty} \hspace{-0.1cm} \sqrt{\frac{\hbar\omega_m}{\ell_{\text{eff}}\mathcal{A}_{r}\epsilon_0}}\Big({a}_{m} - {a}^{\dagger}_{m}\Big)\cos{\Big [\frac{\omega_m}{c}(x+\ell_{c})\Big]},
\end{align}
\end{subequations}
where ${a}_{m}$/$a^{\dagger}_{m}$ is the discrete annihilation/creation operator describing the perfect cavity mode with the resonance frequency $\omega_{m}={\pi c m}/{\ell_{\text{eff}}}$. These operators satisfy the following relations:
\begin{subequations}
\begin{align}
\left[ a_{m},a^{\dagger}_{m'}\right]&=\delta_{mm'},\\
\left[ a_{m},a_{m'}\right]&=0.
\end{align}
\end{subequations}
$\mathcal{A}_{r}$ is the transverse  area of the mode of the perfect cavity, which is related to the mode area of the multilayer cavity with the following relation: $\mathcal{A}_{r}=\left({L_{N}}/{\ell_{\text{eff}}}\right)\mathcal{A}$. Taking this into account the fields in Eq.~\eqref{eq:modes_inside} become 
\begin{subequations}
\label{eq:electric_inside}
\begin{align}
\mathbb{E}_{\text{ins}}(x)&=\sum_{m=1}^{\infty}\hspace{-0.05cm} \sqrt{\frac{\hbar\omega_m}{L_{N}\mathcal{A}\epsilon_0}}\Big({a}_{m} + {a}^{\dagger}_{m}\Big)\sin{\Big [\frac{\omega_m}{c}(x+\ell_{c})\Big]},\\
\mathbb{B}_{\text{ins}}(x)&=\frac{i}{c}\sum_{m=1}^{\infty}  \hspace{-0.05cm}\sqrt{\frac{\hbar\omega_m}{L_{N}\mathcal{A}\epsilon_0}}\Big({a}_{m} - {a}^{\dagger}_{m}\Big)\cos{\Big [\frac{\omega_m}{c}(x+\ell_{c})\Big]}.
\end{align}
\end{subequations}
Analogously, we can define the fields for the outside:
\begin{subequations}
\begin{align}
\mathbb{E}_{\text{outs}}(x)&=\int^{\infty}_{0} \hspace{-0.15cm}d\omega \,\sqrt{\frac{\hbar \omega}{\pi c \mathcal{A}\epsilon_{0}}}\Big(b_{\omega}+b_{\omega}^{\dagger}\Big)\sin\Big[\frac{\omega}{c}(x-d)\Big],
\\
\mathbb{B}_{\text{outs}}(x)&=\frac{i}{c}\int_{0}^{\infty}\hspace{-0.15cm}d\omega\,\sqrt{\frac{\hbar\omega}{\pi c\mathcal{A}\epsilon_{0}}}\Big(b_{\omega}-b_{\omega}^{\dagger}\Big)\cos\Big[\frac{\omega}{c}(x-d)\Big],
\end{align}
\end{subequations}
which are derived considering that the outside is delimited by a perfect mirror on the left, placed at  $d=\ell_{\text{eff}}-\ell_{c}$. Here, unlike for the inside modes, the mode area  $\mathcal{A}$ for the outside is the same as the one for the multilayer cavity. ${b}_{\omega}$/$b^{\dagger}_{\omega}$ is the annihilation/creation operator of the outside, i.e., the reservoir, satisfying the commutation relations
\begin{subequations}
\begin{align}
\left[ {b}_{\omega}, {b}^{\dagger}_{\omega'}\right]&=\delta(\omega - \omega'),\\
\left[ {b}_{\omega}, {b}_{\omega'}\right]&=0.
\end{align}
\end{subequations}
 We highlight that the modes derived here are obtained for a fixed single mode $m$, enforcing a node at distance $d$. {The restriction to a single mode can be justified by the fact that, when an emitter is introduced into the cavity—under the conditions of high finesse and on-resonance coupling —the emitter couples to a single cavity mode, effectively post-selecting the mode of interest, $\omega_m$ (and therefore $\ell_{\text{eff}}$)}. In general, since $\ell^{(m)}_{\text{eff}}$ is different for each $m$, this implies that depending on the mode we study, the position of the replaced mirror $d: = d^{(m)}$ will be different. As an example, for $N = 8$ and $\ell_c = 1.75\ell_0$, the dominant resonance lies at $\omega \approx 1.065 \omega_c$, whereas the geometric cavity length would predict $\omega = 2\pi c/{\ell_c} \approx 1.143 \omega_c$, i.e., a shift of about $7\%$, corresponding to about nine cavity linewidths. A model based on the geometric length would therefore misidentify the resonance coupled to the emitter~\cite{multilayer}.
 
 \section{Cavity - reservoir coupling function \label{sec:coupling}}
 
{If the separation into discrete cavity modes (inside) and the semi-infinite reservoir modes (outside) is correct, then the modes in this representation should be equivalent to those obtained from the true-mode representation. In the following, we impose such equivalence and analyze its validity via deriving the corresponding cavity-reservoir coupling function.} In order to obtain the relation between the true modes ${a}_{\omega}$ and the inside-outside modes ${a}_{m}$ and ${b}_{\omega}$, we equate the corresponding electric and magnetic fields of the inside and the outside parts:
 \begin{subequations}
 \begin{align}
 \mathbb{E}_{\text{ins}}(x)&={E}_{\text{ins}}(x),\\
\mathbb{B}_{\text{ins}}(x)&={B}_{\text{ins}}(x),\\
 \mathbb{E}_{\text{outs}}(x)&={E}_{\text{outs}}(x),\\
\mathbb{B}_{\text{outs}}(x)&={B}_{\text{outs}}(x).
 \end{align} 
 \end{subequations}
 From these equivalences, for the cavity operator $a_{m}$ we obtain (see the details in Appendix~\ref{app:in_operator})
 \begin{align}
 {a}_{m}=&\,  \frac{1}{\ell_{\text{eff}}}\sqrt{\frac{cL_{N}}{\pi}}\int^{\infty}_{0} \hspace{-0.15cm} d\omega\sqrt{\frac{\omega}{\omega_{m}}}\hspace{-0.1cm} \left(\frac{\sin\left[\left(\omega-\omega_{m}\right)\frac{\ell_{\text{eff}}}{c}\right]}{\omega-\omega_{m}}e^{i\frac{\omega}{c}\ell_{c}}\mathcal{T}_{\omega}{a}_{\omega}-\frac{\sin\left[\left(\omega+\omega_{m}\right)\frac{\ell_{\text{eff}}}{c}\right]}{\omega+\omega_{m}}e^{-i\frac{\omega}{c}\ell_{c}}\mathcal{T^{\ast}}_{\omega}{a}_{\omega}^{\dagger}\right), 
 \end{align}
 and for the reservoir operator we get (see the details in Appendix~\ref{app:out_operator}):
 \begin{align}
 \label{eq:b_integral}
 {b}_{\omega}=&\, \frac{1}{2\pi c}\int_{d}^{\infty}dx\int_{0}^{\infty}d\omega'\sqrt{\frac{\omega'}{\omega}} \Bigg(\left[G_{\omega'}e^{i\frac{\omega'-\omega}{c}\left(x-d\right)}+ e^{-i\frac{\omega'}{c}d}e^{-i\frac{\omega'-\omega}{c}\left(x-d\right)}\right]a_{\omega'}\\ \nonumber
 &-\left[G^{\ast}_{\omega'}e^{-i\frac{\omega'+\omega}{c}\left(x-d\right)}+e^{i\frac{\omega'}{c}d}e^{i\frac{\omega'+\omega}{c}\left(x-d\right)}\right]a_{\omega'}^{\dagger}\Bigg),
 \end{align}
 where {for readability, we have introduced the following phase factor:}
 \begin{align}
 \label{eq:R_om}
 G_{\omega'}=e^{2i\frac{\omega'}{c}\ell_{c}}\dfrac{\mathcal{T}_{\omega'}}{\mathcal{T}_{\omega'}^{\ast}}e^{i\frac{\omega'}{c}d}.
 \end{align}
Applying the limit $\mathcal{T}_{\omega}\approx \mathcal{T}_{m}(\omega)\approx T_{m,\omega}$, $G_{\omega}$ simplifies to (see the details in Appendix~\ref{app:out_operator})
 \begin{eqnarray}
 \label{eq:Rom_simp}
 G_{\omega'}\approx e^{-i\frac{\omega'}{c}d}+ 2ie^{i\frac{\omega'}{c}\ell_{c}}{T}_{m,\omega'}\sqrt{\frac{L_N}{\ell_{\text{eff}}}} \, \frac{\cos\left(\frac{\omega'}{c}\ell_{\text{eff}}+\varphi_{r}(\omega')\right)+|r_{m}(\omega')|\cos\left(\frac{\omega'}{c}\ell_{\text{eff}}\right)}{|t_{m}(\omega')|}.
 \end{eqnarray}
 Using this simplification the expression in Eq.~\eqref{eq:b_integral} becomes [see Appendix~\ref{app:out_operator}]:
 \begin{align}
 \nonumber
 {b}_{\omega}	=&\;	e^{-i\frac{\omega}{c}d}{a}_{\omega} +\,\frac{1}{\pi}\int_{0}^{\infty}d\omega'\;\sqrt{\frac{\omega'}{\omega}} \sqrt{\frac{L_{N}}{\ell_{\text{eff}}}}\, \frac{\cos\left(\frac{\omega'}{c}\ell_{\text{eff}}+\varphi_{r}(\omega')\right)+|r_{m}(\omega')|\cos\left(\frac{\omega'}{c}\ell_{\text{eff}}\right)}{|t_{m}(\omega')|}\\ 
 &\times\, \left(e^{i\frac{\omega'}{c}\ell_{c}}{T}_{m, \omega'}\lim_{\epsilon\rightarrow0}\frac{1}{\omega-\omega'-i\epsilon}{a}_{\omega'}+ e^{-i\frac{\omega'}{c}\ell_{c}}{T}_{m,\omega'}^{\ast}\lim_{\epsilon\rightarrow0}\frac{1}{\omega+\omega'-i\epsilon}{a}_{\omega'}^{\dagger}\right).
 \end{align}

 \subsection{Separation of the true modes into inside-outside modes}
 
Above, we derived the expressions of the inside and the outside mode operators in terms of the true modes $a_{\omega}$. Here, we derive the inverse relation assuming that the true modes can be decomposed in terms of the discrete cavity modes and the continuous reservoir modes as follows:
 \begin{equation}
 \label{eq:decomposition_in_out}
 \begin{split}
{a}_{\omega}=&\sum_{m=1}^{\infty}\left [ \alpha_{m_{{-}}}(\omega){a}_{m}+\alpha_{m_{{+}}}(\omega){a}^{\dag}_{m}\right ]+\int_0^{\infty} d\omega' \left (\beta_{{-}}(\omega, \omega') {b}_{\omega'}+\beta_{{+}}(\omega, \omega') {b}^{\dag}_{\omega'}\right),
\end{split}
\end{equation}
{where coefficients $\alpha_{m_{-}}(\omega)$, $\alpha_{m_{+}}(\omega)$ and $\beta_{-}(\omega, \omega')$, $\beta_{+}(\omega, \omega')$ are frequency dependent functions relating the true-modes to the separated inside-outside modes. Below we derive the explicit forms of these coefficients.} Taking the discrete and continuous parts of this definition separately, we further define the following operators: 
 \begin{subequations}
 \label{eq:separation_of_mode_disc_cont}
\begin{align}
{a}_D(\omega)&=\sum_{m=1}^{\infty} \left[\alpha_{m_{{-}}}(\omega){a}_{m}+\alpha_{m_{{+}}}(\omega){a}^{\dag}_{m}\right],\\
{a}_C(\omega)&=\int_0^{\infty} \hspace{-0.25cm}d\omega' \hspace{-0.1cm} \left(\beta_{{-}}(\omega, \omega') {b}_{\omega'}+\beta_{{+}}(\omega, \omega') {b}^{\dag}_{\omega'}\right)
\end{align}
 \end{subequations}
such that 
\begin{align}
{a}_{\omega}={a}_D(\omega)+{a}_C(\omega).  
\end{align}
If the decomposition~\eqref{eq:decomposition_in_out} is indeed accurate, then the following commutation relation must hold true: 
\begin{eqnarray}
\label{eq:separated_com}
\left[{a}_{\omega},{a}^{\dagger}_{\omega'}\right]=\left[{a}_D(\omega),{a}^{\dagger}_D(\omega')\right]+\left[{a}_C(\omega),{a}^{\dagger}_C(\omega')\right]\hspace{-0.1cm}.
\end{eqnarray}
 In order to check this relation we first calculate the coefficients $\alpha_{m_{{{-}}}}(\omega),\, \alpha_{m_{{+}}}(\omega)$ and $\beta_{{{-}}}(\omega,\omega'),\, \beta_{{{+}}}(\omega,\omega')$:
  \begin{subequations}
  \label{eq:discrete_with_global}
 \begin{align}
 \label{eq:alpha1full}
 \alpha_{m_{{-}}} (\omega)&=[{a}_{\omega},{a}^{\dag}_{m}]=\sqrt{\frac{ L_{N}}{\pi c}}\sqrt{\frac{\omega}{\omega_m}} {\text{sinc}\big[{( \omega-\omega_m)\frac{\ell_{\text{eff}}}{c}}\big]}e^{-i\frac{\omega}{c}\ell_c}\mathcal{T}^{\ast}_{\omega},\\
\alpha_{m_{{+}}} (\omega)&=[{a}_{m},{a}_{\omega}] = \sqrt{\frac{ L_{N}}{\pi c}}\sqrt{\frac{\omega}{\omega_m}} {\text{sinc}\big[{( \omega+\omega_m)\frac{\ell_{\text{eff}}}{c}}\big]}e^{-i\frac{\omega}{c}\ell_c}\mathcal{T}^{\ast}_{\omega},
 \end{align}
  \end{subequations}
   \begin{subequations}
   \label{eq:continious_with_global}
 \begin{align}
 \beta_{{{-}}}(\omega,\omega')&=\left[{a}_{\omega},{b}_{\omega'}^{\dagger}\right] =e^{i\frac{\omega'}{c}d}\delta(\omega-\omega')+B(\omega)\sqrt{\frac{\omega}{\omega'}}\lim_{\epsilon\rightarrow0}\frac{1}{\omega'-\omega+i\epsilon},\\ 
\beta_{{{+}}}(\omega,\omega')&=\left[{b}_{\omega'},{a}_{\omega}\right]=-B(\omega)\sqrt{\frac{\omega}{\omega'}}\hspace{-0.1cm}\lim_{\epsilon\rightarrow0}\frac{1}{\omega'+\omega-i\epsilon},
 \end{align}
  \end{subequations}
  where
  \begin{align}
  B(\omega) = \frac{1}{\pi} e^{-i\frac{\omega}{c}\ell_{c}}{T}^{\ast}_{m,\omega}\sqrt{\frac{L_{N}}{\ell_{\text{eff}}}} \frac{\cos\left(\frac{\omega}{c}\ell_{\text{eff}}+\varphi_{r}(\omega)\right)+|r_{m}(\omega)|\cos\left(\frac{\omega}{c}\ell_{\text{eff}}\right)}{|t_{m}(\omega)|}.
  \end{align} 
 In general, operators~\eqref{eq:separation_of_mode_disc_cont} with coefficients~\eqref{eq:discrete_with_global} and~\eqref{eq:continious_with_global} do not satisfy the commutation relation~\eqref{eq:separated_com}, however, as we show below, since we are in the limit where $\mathcal{T}_{\omega}\approx \mathcal{T}_{m}(\omega)\approx T_{m,\omega}$, we can approximate these coefficients such that they satisfy the commutation relations, therefore justifying the separation of the modes into inside and outside. To demonstrate that, we proceed by approximating coefficients $\alpha_{m_{i}}(\omega)$ and $\beta_{i}(\omega,\omega')$:
 \begin{subequations}
 \label{eq:approximate_coef}
 \begin{align}
 \label{eq:alpha_approx}
 \alpha_{m_{\color{black}{-}}}(\omega)\approx& \; e^{-i\frac{\omega}{c}\ell_c} \sqrt{\frac{{\kappa}_{m}}{2\pi}} \frac{1}{\left( \omega - \omega_{m}\right)-i\frac{{\kappa}_{m}}{2}},\\
 \alpha_{m_{\color{black}{+}}}(\omega)\approx& \; 0,\\
 \nonumber
  \beta_{{\color{black}{-}}}(\omega,\omega')\approx &\; e^{i\frac{\omega'}{c}d}\delta(\omega-\omega')+\frac{1}{\pi} \Bigg(e^{-i\frac{\omega}{c}\ell_{c}}{T}^{\ast}_{m,\omega}\sqrt{\frac{L_{N}}{\ell_{\text{eff}}}}\lim_{\epsilon\rightarrow0}\frac{1}{\omega'-\omega+i\epsilon} (-1)^{m}\frac{-1+|r_{m}(\omega_{m})|}{|t_{m}(\omega_{m})|}\Bigg),\\
 \beta_{{\color{black}{+}}}(\omega,\omega')\approx &\;0,
  \end{align}
  \end{subequations}
 where we have  evaluated $\alpha_{m_{\color{black}{-}}}(\omega)$ and  $ \beta_{{\color{black}{-}}}(\omega,\omega')$ around the resonance peak $\omega_{m}$ and discarded $\alpha_{m_{\color{black}{+}}}$ and $\beta_{{\color{black}{+}}}(\omega,\omega')$ treating them as fast oscillating terms. 
 These approximations lead to the following relation between the true and the separated inside-outside modes:
 \begin{equation}
 \label{eq:mode_relation_global_separate}
{a}_{\omega}\approx \sum_{m=1}^{\infty}\alpha_{m_{\color{black}{-}}}(\omega){a}_{m}+\int_0^{\infty} d\omega' \beta_{\color{black}{-}}(\omega, \omega') {b}_{\omega'}.
\end{equation}
 With this separation the commutation relation~\eqref{eq:separated_com} is satisfied, {hence the presented separation of the modes into inside and outside ones is accurate} as long as {$|\omega-\omega_m|\ll {\kappa}_m$ and} the terms of the order $\left( {\kappa}_{m} \frac{\ell_{\rm{eff}}}{c}\right)$ are negligible, i.e., the reflectivity of the mirror is close to 1 [Fig.~\ref{fig:fig2}(b)] (see the details in Appendix~\ref{app:commutation_satisfy}).

  \subsection{Hamiltonian}
  
  Having obtained the relation between the modes of the true-mode and the inside-outside representations,  from the true-mode Hamiltonian~\eqref{eq:global_Hamiltonian} we can derive the Hamiltonian describing the separated inside-outside system (see the details in Appendix~\ref{app:Hamiltonian}):
 \begin{eqnarray}
H = \sum_{m=1}\hbar \omega_{m}{a}^{\dagger}_{m}{a}_{m}+\int_{0}^{\infty}d\omega\:\hbar\omega{b}_{\omega}^{\dagger}{b}_{\omega}+ i\hbar \sum_{m=1}\int^{\infty}_{0} d\omega \, \left({V}_{m}(\omega){b}^{\dagger}_{\omega}{a}_{m}- {V}_{m}^{\ast}(\omega){a}^{\dagger}_{m}{b}_{\omega}\right),
\end{eqnarray} 
with
 \begin{align}
 \label{eq:cavity_res_coupling}
 {V}_{m}(\omega) = -ie^{-i\frac{\omega}{c}\ell_{\rm{eff}}}\sqrt{\frac{\kappa_{m}}{2\pi}} \, \text{sinc}\left[ (\omega-\omega_{m})\frac{\ell_{\rm{eff}}}{c}\right]
 \end{align}
 being the coupling between the cavity and the environment. As we can see, in contrast to the phenomenological model, where the outside is treated as a constant loss accounted for via a constant decay rate $\kappa_m$, the latter is a function of the characteristic parameters of the cavity as well as the light frequency in the inside-outside model. Furthermore, unlike in phenomenological models, $\kappa_m$ also varies with the cavity mode number $m$ due to the multilayered nature of the cavity mirror. We observe that, unlike the cavity-environment coupling function derived for a cavity with a single-layered mirror of negligible thickness~\cite{dutra2005cavity, Saharyan5.033056}, Eq.~\eqref{eq:cavity_res_coupling} features the effective length $\ell_{\rm{eff}}$ and the linewidth ${\kappa}_m$ of the multilayer cavity. Just like the response function~\eqref{eq:resp_sing_for_mult}, the coupling function~\eqref{eq:cavity_res_coupling} can be considered as a generalization of the similar expression derived for a single-layered cavity~\cite{dutra2005cavity}. Only in a certain limit, where $\ell_{\text{eff}}=\ell_{c}$ these formulas coincide. 
 
 \subsection{Domain of validity of the inside-outside model}
 
 The effective inside--outside description derived above relies on several approximations. First, the reduction of the full multilayer response to a single Lorentzian contribution requires sufficiently isolated resonances, characterized by the condition $\zeta_m\ll 1$ and $\kappa_m\ell_{\mathrm{eff}}/c\ll1$, and a reflection magnitude and round-trip phase that can be approximated as constant and linear, respectively, across the selected linewidth. These conditions are generally favored by narrow resonances of highly reflecting multilayer mirrors, featuring large $N$, although high reflectivity alone does not guarantee negligible spectral overlap. The subsequent separation into discrete cavity and continuous reservoir modes is performed around the selected resonance, where the frequency-dependent coefficients are approximated in the narrow spectral region $|\omega-\omega_m|\ll\kappa_m$, and counter-rotating terms are neglected. Furthermore, as demonstrated in Appendix D, the approximate preservation of the canonical commutation relations requires $\kappa_m\ell_{\mathrm{eff}}/c\ll1$. Finally, the optical description assumes a single paraxial transverse mode, with transverse effects incorporated through the effective mode area $\mathcal A$, neglecting the angular and polarization dependence of the multilayer response. Outside these regimes, the full true-mode description remains applicable within the one-dimensional model, while a more general multimode or transverse-mode treatment may be required.

 \subsection{Poynting vector}
 
 Now that we have a complete description of the model in inside-outside representation, here we focus on  characterizing the field propagating outside the cavity. In the context of quantum networking between two spatially separated quantum nodes, the absorption of a photon by the receiving node can be regarded as a single-photon detection process. {During the photon detection process, the likelihood for a photon detection at time $t$ is proportional to the probability density of photons at the position of the detector at time $t$. However the description of the process of detection would require a model of the measurement apparatus itself. The construction of such model is beyond the scope of this paper, but it would require the precise characterization of the photon profile and its propagation, which can be obtained by the Poynting vector ~\cite{dutra2005cavity},} We start from the true-mode representation, where the corresponding Poynting vector for the outside of the cavity reads as follows:
 \begin{align}
S_{\text{outs}}(x)= -\frac{1}{2\mu_{0}}(B_{\text{outs}}(x)E_{\text{outs}}(x)+E_{\text{outs}}(x)B_{\text{outs}}(x)),
\end{align} 
where $\mu_{0}$ is the free space permeability. Using expressions~\eqref{eq:Eout} and~\eqref{eq:Bout}, for the propagation in the positive $x$ direction, the above expression can be written as
\begin{align}
S_{\text{outs}}(x) = \frac{\hbar}{2\pi \mathcal{A}} \hspace{-0.1cm} \int_{0}^{\infty} \hspace{-0.25cm} d\omega d\omega' \sqrt{\omega \omega'} \text{Re}\left \{R_{\omega} R^{\ast}_{\omega'}e^{i \frac{(\omega-\omega')}{c}x}a^{\dagger}_{\omega'}a_{\omega}\right \}
\end{align}
Taking the definition in Eq.~\eqref{eq:R_om} and the expression of the coefficient $\alpha_{m_{{\color{black}{-}}}}(\omega)$ given in Eq.~\eqref{eq:alpha_approx}, we can reformulate $G_{\omega}$ as follows: 
\begin{align}
\label{eq:Rvsalpha}
G_{\omega} \approx e^{2i\frac{\omega}{c}\ell_{c}}\dfrac{\mathcal{T}_{m}(\omega)}{\mathcal{T}^{\ast}_{m}(\omega)}e^{i\frac{\omega}{c}d}\approx e^{i\frac{\omega}{c}\ell_{\text{eff}}}{\alpha^{\ast}_{m_{{\color{black}{-}}}}(\omega)}\sqrt{\frac{2\pi}{\textcolor{black}{\kappa}_{m}}}\left( \omega-\omega_{m}-i\frac{\textcolor{black}{\kappa}_{m}}{2}\right).
\end{align}
Using Eq.~\eqref{eq:Rvsalpha} and the expansion of true modes in terms of outside modes: $a_{\omega} = \int_{0}^{\infty}d\omega'' \beta_{{\color{black}{-}}} (\omega,\omega'') b_{\omega''}$, we can write the Poynting vector for the outgoing field in the inside-outside representation:
\begin{align}
\label{eq:Poynting_out}
{S}_{\text{outs}}(x) \approx \frac{\hbar \omega_{m}}{\mathcal{A} \textcolor{black}{\kappa}_{m}}\int_{0}^{\infty}d\omega\, d\omega' \; \textcolor{black}{V}_{m}^{\ast}(\omega) \textcolor{black}{V}_{m} (\omega')e^{i \frac{(\omega-\omega')}{c}x}b^{\dagger}_{\omega'}b_{\omega},
\end{align}
where we have evaluated the integral
\begin{align*}
 \int^{\infty}_{0} d\omega'' \hspace{-0.1cm}\left(  \omega-\omega_{m}-i\frac{\textcolor{black}{\kappa}_{m}}{2} \right) \alpha_{m_{{\color{black}{-}}}}^{\ast}(\omega) \beta_{{\color{black}{-}}}(\omega,\omega'')b_{\omega''}  =  -i \textcolor{black}{V}_{m}^{\ast}(\omega)b_{\omega}
 \end{align*}
similar to the one calculated in Appendix~\ref{app:Hamiltonian}. 

Having obtained the expression for the Poynting vector, we next define the following operator that characterizes the spatial distribution of the photon propagating outside~\cite{Saharyan5.033056}: 
\begin{eqnarray}
\label{eq:bx}
b(x) :=  \frac{1}{\sqrt{\textcolor{black}{\kappa}_{m}}}\int^{\infty}_{0} d\omega \textcolor{black}{V}_{m}^{\ast}(\omega) e^{i\frac{\omega}{c}x}b_{\omega},
\end{eqnarray}
{which is different from the standard Fourier transform definition by the fact that it has the coupling function $\textcolor{black}{V}_m (\omega)$ in the expression. While qualitatively this is not very different from the Fourier transform, given that $|\textcolor{black}{V}_m(\omega_m)|\approx \sqrt{\textcolor{black}{\kappa}_m/2\pi}$, introduction of such a $\textcolor{black}{V}_m(\omega)$ function allows us to evaluate the integrals in a mathematically more straightforward manner, and this is a definition derived from the first principles.} The Poynting vector can then be written as
\begin{eqnarray}
\label{eq:Souts}
{S}_{\text{outs}}(x) = \frac{\hbar \omega_{m}}{\mathcal{A}}b^{\dagger}(x)b(x), 
\end{eqnarray}
which is the energy flow of the propagating photon at a position $x$. \textcolor{black}{Furthermore, recalling the Poynting theorem~\cite{garrison2008quantum,dutra2005cavity}, for the expectation value of the Poynting vector corresponding to the energy flow of linearly polarized light propagating in one-dimensional space, we can write the following:
 \begin{align}
 \label{eq:theorem_Poynting}
 \frac{\partial}{\partial x}\langle {S}_{\text{outs}}(x)\rangle_{\lvert \psi(t)\rangle} + \frac{\partial}{\partial t} \langle U_{\text{outs}}(x)\rangle_{\lvert \psi(t)\rangle}=0,
 \end{align}
 where $\langle U_{\text{outs}}(x)\rangle_{\lvert \psi(t)\rangle}$ is the expectation value of the electromagnetic energy density. As it is shown in Ref.~\cite{Saharyan5.033056}, if one writes the Poynting vector~\eqref{eq:Souts} in Heisenberg representation, or equivalently the Heisenberg representation of operator $b(x)$ in Eq.~\eqref{eq:bx}, it can be shown that this term depends only on $t-x/c$ (we also confirm this in the following analysis of the dynamics, at the limit where the photon is completely outside the cavity, the definition~\eqref{eq:photon_space_coef}, which is obtained via calculating the expectation value of $b^{\dagger}_{x}b_{x}$, depends only on $t-x/c$). Taking this into account, Eq.~\eqref{eq:theorem_Poynting} can be equivalently written as 
 \begin{align}
 -\frac{\partial}{c\partial t}\langle {S}_{\text{outs}}(x)\rangle_{\lvert \psi(t)\rangle}+ \frac{\partial}{\partial t} \langle U_{\text{outs}}(x)\rangle_{\lvert \psi(t)\rangle}=0,\\
\color{black}{\frac{\partial}{\partial x}\langle {S}_{\text{outs}}(x)\rangle_{\lvert \psi(t)\rangle}-c \frac{\partial}{\partial x} \langle U_{\text{outs}}(x)\rangle_{\lvert \psi(t)\rangle}=0},
 \end{align}
 leading to the following relation between the Poynting vector and the energy density:
  \begin{align}
  \label{eq:Poynting_energy}
 \frac{\langle {S}_{\text{outs}}(x)\rangle_{\lvert \psi(t)\rangle}}{c}= \langle U_{\text{outs}}(x)\rangle_{\lvert \psi(t)\rangle}.
 \end{align}
{{As mentioned at the beginning of the section, the detector measures the probability density of the photon, which, unlike the Poynting vector, is independent of the energy. Thus,} using the fact that $\psi(t)$ is a quasi-monochromatic state, one can define an averaged coarse-grained photon number density  propagating through the area $\mathcal{A}$ at a position $x$ by 
\begin{align}
\Phi_{\text{density}}(x,t) := \frac{\mathcal{A}}{\hbar \omega_{m}}{\langle {U}_{\text{outs}}(x)\rangle_{\lvert \psi(t)\rangle}},
\end{align}
which using~\eqref{eq:Poynting_energy} can be expressed as}
\begin{align}
\label{eq:density}
 \Phi_{\text{density}}(x,t)  &= \frac{\mathcal{A}}{\hbar \omega_{m}}\frac{\langle {S}_{\text{outs}}(x)\rangle_{\lvert \psi(t)\rangle}}{c}=\frac{1}{c}\langle b^{\dagger}(x)b(x)\rangle=\langle b^{\dagger}_{x} b_{x} \rangle,
\end{align}
where we have defined the operator 
\begin{align}
\label{eq:singlephoton_bx}
b^{\dagger}_{x} = \frac{1}{\sqrt{c}} b^{\dagger}(x)= \frac{1}{\sqrt{c\textcolor{black}{\kappa}_{m}}}\int^{\infty}_{0} d\omega \textcolor{black}{V}_{m}(\omega) e^{-i\frac{\omega}{c}x}b^{\dagger}_{\omega},
\end{align}
which satisfies the following commutation relations (see Appendix~\ref{app:photon_in_space}): 
\begin{align}
\left[ b_{x}, b^{\dagger}_{x}\right] &= \frac{1}{2\ell_{\text{eff}}},\\
\left[ b_{x}, b^{\dagger}_{x'}\right] &= 0, \quad \text{when} \; |x-x'|>2\ell_{\text{eff}}.
\end{align}
Taking this definition into account, the total photon number outside the cavity ($x>0$) can be written as follows: 
\begin{align}
\label{eq:number_space}
N(t) = \int^{\infty}_{0} dx \, \Big \langle b^{\dagger}_{x} b_{x} \Big \rangle_{\lvert \psi(t)\rangle}.
\end{align}
We remark that this definition is possible due to the fact that we consider a one-dimensional model for the propagating light and limit the study to a single mode $\omega_{m}$. As discussed in literature~\cite{Mandel_1966,garrison2008quantum,Federico_2023_2}, in a more general case, a definition of a local photon number operator or a local energy density is not possible since the process of projecting out the transverse and longitudinal part of a vector field is a nonlocal operation in position space. However in the particular case discussed here, we are able to define such a local operator in a so-called coarse-grained limit, due to the explicit form of $\textcolor{black}{V}_{m}(\omega)$ in the definition~\eqref{eq:singlephoton_bx} (see the discussion in Appendix~\ref{app:photon_in_space}).}

The photon density in Eq.~\eqref{eq:density}, is related to the photon flux derived in Ref.~\cite{Saharyan5.033056} as follows:
\begin{equation}
 \Phi_{\text{flux}}(x,t)  =  c \, \Phi_{\text{density}}(x,t) 
\end{equation} 
Additionally, we can define a general photon state outside the cavity as~\footnote{Eq.~\eqref{eq:1out} gives the general form of the single photon state which satisfies $n=1$ [Eq.~\eqref{eq:one_photon}]. This corrects Eq.~(24) given in Ref.~\cite{Saharyan5.033056}}
\begin{eqnarray}
\label{eq:1out}
\lvert 1_{\text{outs}}(t)\rangle = \int_0^{+\infty}\hspace{-1em}d\omega \;{\Psi}(\omega,t) b^{\dagger}_{\omega} \lvert \emptyset \rangle,
\end{eqnarray}
with the shape ${\Psi}(\omega,t)$, where the time dependence appears in the phase corresponding to the propagation such that at all times $t$
\begin{equation}
\int^{\infty}_{0} d\omega \lvert {\Psi}(\omega,t) \rvert^{2} =1,
\end{equation}
which satisfies 
\begin{align}
\label{eq:one_photon}
N = \int^{\infty}_{0} dx \, \Big \langle b^{\dagger}_{x} b_{x} \Big \rangle_{\lvert 1_{\text{outs}}\rangle} = 1.
\end{align}
We remark that $b^{\dagger}_{x}$ cannot be interpreted as an operator that creates a photon at position $x$; such operator does not exist since the mode of the photon is non-local~\cite{garrison2008quantum,Federico_2023_2}. Nevertheless, when considering the propagating photon as an information carrier between distant nodes, $b(x)$ from Eq.~\eqref{eq:bx} is the one capturing the spatial properties of the outgoing photon.

In the following, we use the definitions presented here to analyze the production and the propagation of a single photon produced from an atom trapped in a cavity. 

 \section{Numerical verification of the inside-outside representation \label{sec:verification}}

 \begin{figure}[!ht]
\begin{center}
\includegraphics[scale=0.55]{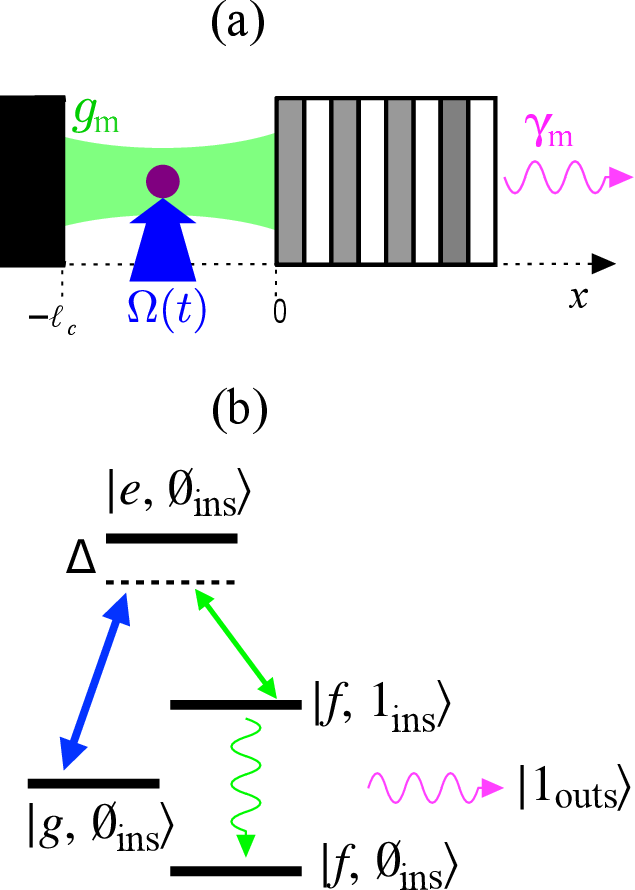}
\caption{ (a) $\Lambda$-type atom trapped in a non-perfect cavity. The cavity is formed by a perfect mirror placed at $x=-\ell_{c}$ and a partially transparent mirror standing from $x=0$ and made of a dielectric stack of alternating layers. The atom is  driven  by  an  external
    classical  laser field  of  Rabi frequency  $\Omega(t)$. $g_{m}$ is the coupling strength between the atom and the $m$-th mode of the cavity, and $\textcolor{black}{\kappa}_{m}$ is the corresponding linewidth. (b) Inside - outside representation of the considered model. Initially the atom  is in
    the ground state $|g\rangle$. In the course of the excitation process, one
  photon is taken from the  laser field and transferred to the cavity, corresponding to the state $\lvert f,1_{\rm{in}}\rangle$. 
  Due to the cavity leakage $\textcolor{black}{\kappa}_{m}$, this photon eventually  leaks   out   of  the   cavity  to the environment transferring the photon state to the outside ($\lvert 1_{\rm{out}}\rangle$).}
\label{fig:scheme}
\end{center}
\end{figure}

 To verify the validity of the above mode separation, we study the dynamics of a $\Lambda$-type atom of ground state $\lvert g \rangle$, excited state $\lvert e \rangle$ and a metastable state $\lvert f \rangle$, trapped in a cavity with a multilayer dielectric mirror (Fig.~\ref{fig:scheme}). The atom is excited by a classical laser field of Rabi frequency $\Omega(t)$, introducing a detuning between the laser frequency $\omega_{L}$ and the frequency of excited state $\omega_{\rm{eg}}$: $\Delta = \omega_{\rm{eg}}-\omega_{L}$. Furthermore, via coupling to the cavity, the atom decays from the excited to the metastable state, emitting a photon that eventually leaks out of the cavity through the partially transparent dielectric mirror. To study the dynamics of this process, we start by analyzing this model in terms of the true modes. For this case, the overall Hamiltonian can be written as:
  \begin{subequations}
\begin{align}
\label{eq:global_Ham}
&\tilde{H}(t)={H}_{A}(t)+{H}_{\text{int}}+{H}_{E}\\
\label{eq:atom_hamiltonian}
&{H}_{A}(t)=\hbar (\omega_{\rm{fg}}-\omega_{L})\sigma_{\rm{ff}}+\hbar \Delta \sigma_{\rm{ee}}+\hbar\Omega (\sigma_{\rm{ge}} +\sigma_{\rm{eg}}),\\
&{H}_{E} = \int_0^{+\infty} d\omega \; \hbar \omega a^{\dagger}_{\omega}a_{\omega},\\
&{H}_{\text{int}} = i\hbar \int_0^{+\infty}  d\omega \; \left( \eta_{\omega} a_{\omega} \sigma_{\rm{fe}}^{\dagger } - \eta^{\ast }_{\omega} a^{\dagger}_{\omega} \sigma_{\rm{fe}} \right),
\end{align}
 \end{subequations}
where the Hamiltonian $H_{E}$ represents the part describing the {``universe''}, which is the multilayer cavity and reservoir considered as a single entity. We have also introduced the atomic operators $\sigma_{ij}=\lvert i \rangle \langle j\rvert$. $\eta_{\omega}$ describes the coupling of the emitter with the true mode $a_{\omega}$ \cite{multilayer}:
\begin{align}
\eta_{\omega} &= i \sqrt{\frac{\omega}{\hbar \epsilon_{0} \pi c \mathcal{A}}} d_{\rm{fe}}e^{i\frac{\omega}{c}\ell_{c}}\sin{\left( \frac{\omega}{c}(x_{A}+\ell_{c})\right)}\mathcal{T}_{\omega},
\end{align}
with $x_{A}$ being the position of the emitter and $d_{\rm{fe}}$ being the dipole moment between levels $\lvert f \rangle \rightarrow \lvert e \rangle $. 

The state corresponding to this representation can be written as
\begin{equation}
\label{eq:global_state}
\lvert \tilde{\psi} \rangle = \tilde{c}_{g,0}(t){\lvert g,\emptyset \rangle}+\tilde{c}_{e,0}(t){\lvert e,\emptyset \rangle} +\hspace{-0.3em}\int_0^{+\infty}\hspace{-1em}d\omega\,\tilde{c}_{f,1}(\omega,t)\lvert f,\mathbf{1_{\omega}} \rangle,
\end{equation}
where 
\begin{equation}
\lvert \mathbf{1_{\omega}} \rangle = a^{\dagger}_{\omega} \lvert \emptyset \rangle.
\end{equation}
The dynamical equations corresponding to the Hamiltonian in~\eqref{eq:global_Ham} with the state~\eqref{eq:global_state} are
 \begin{subequations}
 \begin{align}
i\dot{\tilde{c}}_{g,0}(t) =& \; \Omega(t) \, \tilde{c}_{e,0}(t),\\
i\dot{\tilde{c}}_{e,0}(t)=&\; \Delta \, \tilde{c}_{e,0}(t)+\Omega(t) \, \tilde{c}_{g,0}(t)+  i\int_0^{+\infty}\hspace{-1em}d\omega \; \eta_{\omega} \, \tilde{c}_{f,1}(\omega,t),\\
i\dot{\tilde{c}}_{f,1}(\omega,t) =& \; ( \omega +\omega_{\rm{fg}}-\omega_{L})\tilde{c}_{f,1}(\omega,t)-i \eta^{\ast}_{\omega} \tilde{c}_{e,0}(t).
\end{align}
\label{eq:dyn_true}
 \end{subequations}

 The same system in the inside-outside representation has the following Hamiltonian:
  \begin{subequations}
 \begin{align}
&\hat{H}(t)= {H}_{A}(t)+{H}_{AC}+{H}_{C}+{H}_{RS}+{H}_{R}\\
&{H}_C=\sum_{m}\hbar\omega_m a_{m}^{\dagger}a_{m},\\
&{H}_{AC}=\sum_{m}\hbar g_{m} \big(a_{m}^{\dagger}\sigma_{\rm{fe}} +\sigma_{\rm{fe}}^{\dagger }a_{m}\big),\\
&{H}_R=\int_0^{+\infty}\hspace{-1em} d\omega\,\hbar\omega\,b_{\omega}^{\dagger}b_{\omega},\\
&{H}_{RC}=i\hbar\sum_{m}\int_0^{+\infty}\hspace{-1em} d\omega\,\big( \textcolor{black}{V}_{m}(\omega) b_{\omega}^{\dagger}a_{m}
-\textcolor{black}{V}_{m}^{\ast}(\omega) a_{m}^{\dagger}  b_{\omega}\big),
\end{align}
 \end{subequations}
 where $g_{m}$ is the coupling of the atom and the $m$-th mode of the perfect cavity:
 \begin{align}
 g_{m}=-d_{\rm{fe}}\sqrt{\frac{\omega_m}{\hbar\epsilon_0 L_{N} \mathcal{A}}}\sin{\left[ \frac{\omega_{m}}{c}(x_{A}+\ell_{c})\right]}.
 \end{align}
 The state in this representation can be written as follows: 
\begin{align*}
&\lvert \psi \rangle = c_{g,0 ,0}(t)\lvert g,\emptyset_{\text{ins}},\emptyset_{\text{outs}} \rangle+c_{e,0,0}(t)\lvert e,\emptyset_{\text{ins}},\emptyset_{\text{outs}} \rangle \\
&+ \sum_{m}\hspace{-0.1cm}c^{(m)}_{f,1,0}(t)\lvert f,1^{(m)}_{\text{ins}}\hspace{-0.05cm},\hspace{-0.05cm}\emptyset_{\text{outs}} \rangle \hspace{-0.1cm}+\hspace{-0.1cm} \int_0^{+\infty} \hspace{-0.5cm} d\omega \, c_{f,0,1}(\omega,t)\lvert f, \emptyset_{\text{ins}},1_{\text{outs},\omega} \rangle,
\end{align*} 
with 
\begin{subequations}
 \begin{eqnarray}
 \lvert 1_{\rm{in}}^{(m)}\rangle &=& a^{\dagger}_{m}\lvert \emptyset \rangle,\\
 \lvert 1_{\omega ,\rm{out}} \rangle &=& b^{\dagger}_{\omega} \lvert \emptyset_{\rm{out}} \rangle.
 \end{eqnarray}
 \end{subequations}
 The corresponding dynamical equations are
  \begin{subequations}
  \label{eq:dyn_in_out}
 \begin{align}
i\dot{c}_{g,0,0}(t) =&\; \Omega(t) \,c_{e,0,0}(t),\\ 
i\dot{c}_{e,0,0}(t) =&\; \Delta \, c_{e,0,0}(t)+\Omega(t) \, c_{g,0,0}(t)+\sum_{m} g_{m} \, c^{(m)}_{f,1,0}(t),\\ 
i\dot{c}^{(m)}_{f,1,0}(t) =&\;(\omega_{\rm{fg}}+\omega_{m}-\omega_{L})c^{(m)}_{f,1,0}(t) +g_{m} \, c_{e,0,0}(t)+\int_0^{+\infty}\hspace{-1em}d\omega \; \textcolor{black}{V}_{m}^{\ast}(\omega) \, c_{f,0,1}(\omega,t),\\ 
i\dot{c}_{f,0,1}(\omega,t) =&\; \left( \omega_{\rm{fg}} - \omega_{L}+\omega \right)c_{f,0,1}(\omega,t)+\sum_{m}\textcolor{black}{V}_{m}(\omega) \, c^{(m)}_{f,1,0}(t).
\end{align}
 \end{subequations}
 \textcolor{black}{In a regime where the cavity is such that after some time the photon leaks out of it completely, the outside photon state becomes $c_{f,0,1}(\omega,t) \rightarrow \mathbf{c}_{f,0,1}(\omega)e^{-i(\omega_{\rm{fg}} - \omega_{L}+\omega)t}$, where 
 \begin{equation}
 \int^{\infty}_{0} d\omega \lvert {c}_{f,0,1}(\omega,t)\rvert^{2}= \int^{\infty}_{0} d\omega \lvert \mathbf{c}_{f,0,1}(\omega)\rvert^{2} = 1.
 \end{equation}
Thus, at this limit we can identify the function $\Psi(\omega,t)$ in Eq.~\eqref{eq:1out} as 
\begin{equation}
\Psi(\omega,t) = \mathbf{c}_{f,0,1}(\omega)e^{-i(\omega_{\text{fg}} - \omega_{L}+\omega)t}.
\end{equation}
Following the definition of the operator~\eqref{eq:singlephoton_bx} and taking the solutions of Eq.~\eqref{eq:dyn_in_out}, we can determine the density of photon number at a position $x$ by calculating the mean value in Eq.~\eqref{eq:density}:
\begin{align}
\Phi_{\text{density}}(x,t) = \lvert c_{0,1}(x,t)\rvert^{2} = \langle  \psi(t) \rvert b^{\dagger}_{x} b_{x} \lvert \psi (t)\rangle
\end{align}
with
\begin{eqnarray}
\label{eq:photon_space_coef}
c_{0,1}(x,t) =  \frac{1}{\sqrt{c\textcolor{black}{\kappa}_{m}}}\int^{\infty}_{0} d\omega \textcolor{black}{V}_{m}^{\ast}(\omega) e^{i\frac{\omega}{c}x} c_{f,0,1}(\omega,t).
\end{eqnarray}
}
\begin{figure}[!h]
\centering
\includegraphics[scale=0.5]{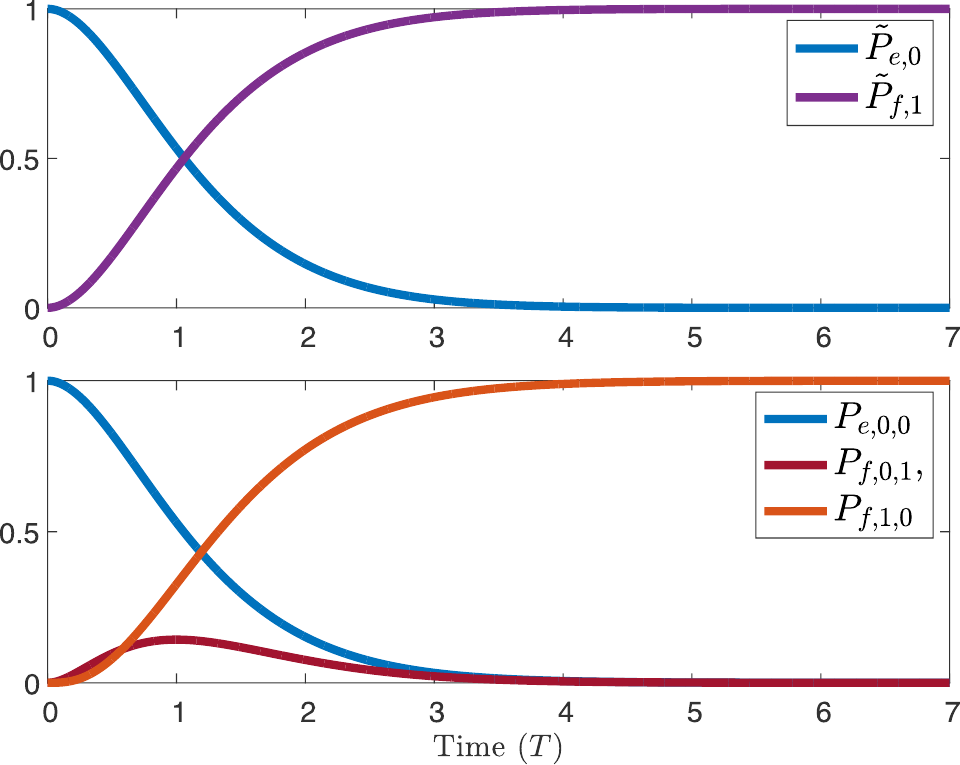}
\caption{ The dynamics in true-mode (top figure) and inside-outside representations (bottom figure). The figures are obtained for the case where the atom is initially in the excited state in a cavity having a multilayer mirror with $N=8$ dielectric layer pairs and the mirror spacing $\ell_{c} = 10\ell_{0}$. We consider the following parameters $\Omega(t) = 0, \, \Delta = 0, \, {d_{\rm{fe}}}/{\sqrt{\hbar \epsilon_{0} c \mathcal{A}}} = 0.0025$ and $x_{A} = -4.5\ell_{0}$. The lines correspond to the coefficients $\tilde{P}_{e,0} = |\tilde{c}_{e,0}(t)|^{2}$, $\tilde{P}_{f,1} = \int_{0}^{\infty} d\omega \, |\tilde{c}_{f,1}(\omega,t)|^{2}$ in the true-mode representation and $P_{e,0,0} = |c_{e,0,0}(t)|^{2}$, $P_{f,1,0} = |c_{f,1,0}(t)|^{2}$ and $P_{f,0,1} = \int_{0}^{\infty} d\omega \, |c_{f,0,1}(\omega,t)|^{2}$ in the inside-outside representation.}
\label{fig:photon_shape}
\end{figure}

\begin{figure}[!h]
\centering
\includegraphics[scale=0.5]{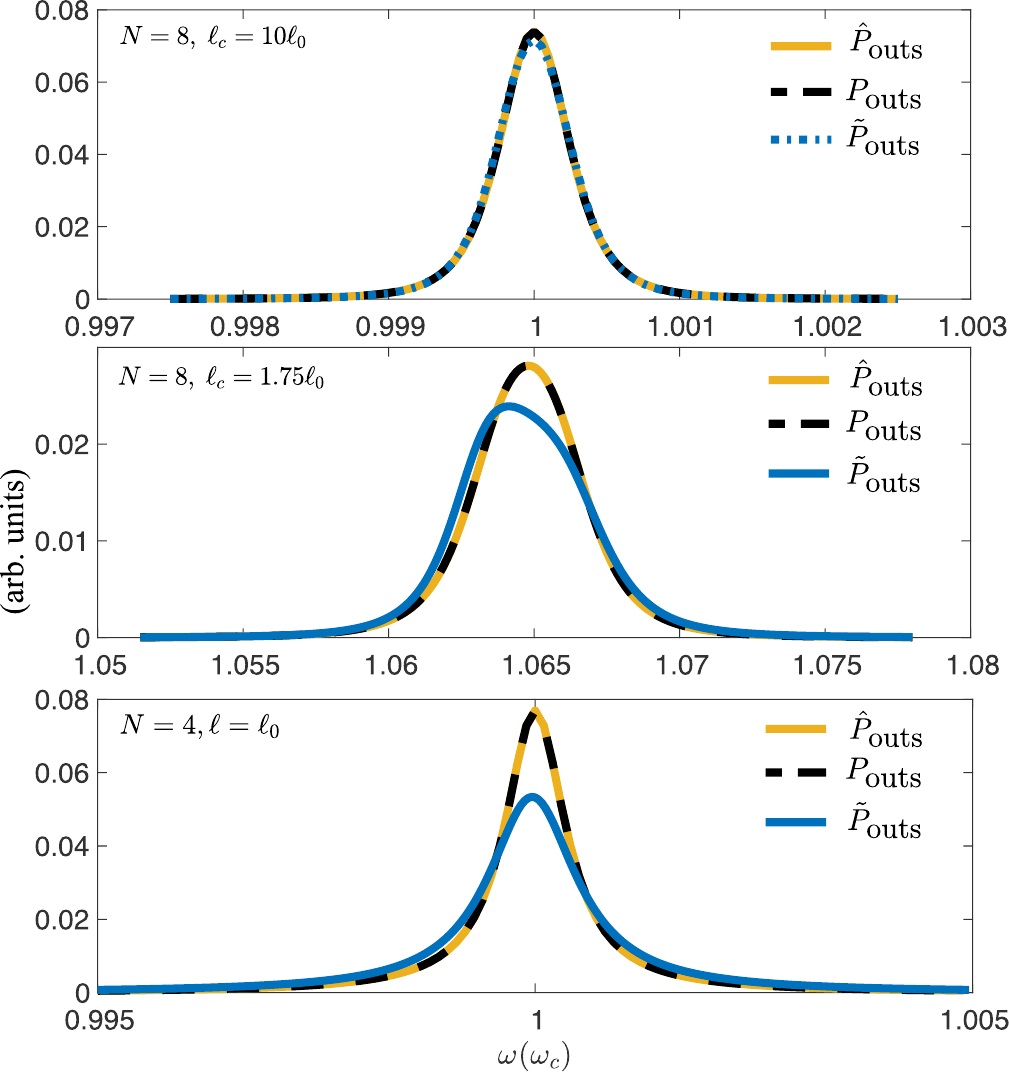}
\caption{The shape of the outgoing photon in the frequency domain, obtained from true-mode representation ($\tilde{P}_\text{outs}$), inside-outside representation (${P}_\text{outs}$) and via a Fourier transform of the inside photon state ($\hat{P}_\text{outs}$). The parameters are defined as follows: $\tilde{P}_{\text{outs}} =| \tilde{c}_{f,1}(\omega,t_{\text{f}})|^2$,  ${P}_\text{outs}=|c_{f,0,1}(\omega,t_f)|^2$, where $t_{\rm{f}}$ is a time corresponding to the steady state of the photon, i.e., the time when the photon has leaked out of the cavity, and $\hat{P}_\text{outs} = \lvert \sqrt{{\textcolor{black}{\kappa}_m}/{2\pi}}\text{FT}\{ c_{f,1,0}(t)\}\rvert^2$. The top figure corresponds to the case described in Fig.~\ref{fig:photon_shape}, with an effective reflectivity $|r_m|\approx 0.953$. In the middle figure, the effective reflectivity is $|r_m|\approx 0.949$ with $\ell_{c}=1.75\ell_{0}$, and $x_{A} \approx -0.342\ell_{0}$, such that the atom of transition frequency $\omega \approx 1.065 \omega_c$ is placed at the field maximum. In the bottom figure $N = 4$, $\ell_{c}=\ell_{0}$ and $x_{A} = -0.5\ell_{0}$, making the effective reflectivity $|r_m|\approx 0.858$. The transition frequency of the atom is $\omega = \omega_c$ with ${d_{\rm{fe}}}/{\sqrt{\hbar \epsilon_{0} c \mathcal{A}}} = 0.01$.}
\label{fig:spectral_photon}
\end{figure}

\begin{figure}[!h]
\centering
\includegraphics[scale=0.45]{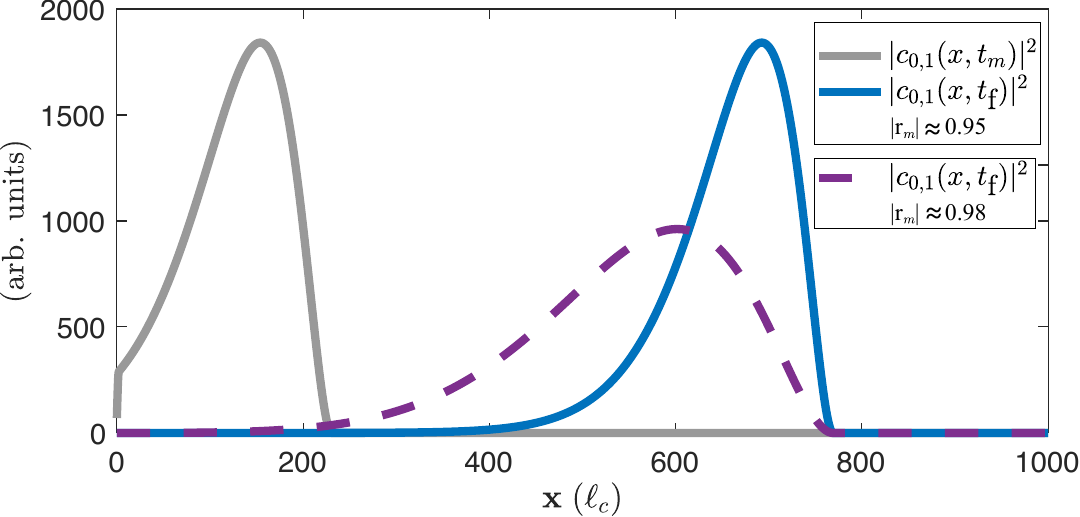}
\caption{Propagating single-photon state along the propagation direction. The cavity mirror of the inside-outside representation is placed at $d=0$ (for the actual multilayer mirror $\ell_{c}=10\ell_{0}$, thus $\ell_{\text{eff}}=\ell_{c}$). The solid curves are obtained with the same parameters as the ones in Fig.~\ref{fig:photon_shape}(a), leading to effective reflectivity $|r_{m}| \approx 0.95$. The curve corresponding to $|c_{0,1}(x,t_{m})|^{2}$ describes the photon state at a time $t_{m}$, when the photon is not completely out of the cavity yet (in the simulation the value of $t_{m}$ corresponds to the point $t_{m} = 3$ in Fig.~\ref{fig:photon_shape}(a)). Similarly, $|c_{0,1}(x,t_{\text{f}})|^{2}$ corresponds to the time $t_{\text{f}}$ when the photon has left the cavity and propagates along the positive $x$ direction. The dashed line is $|c_{0,1}(x,t_{\text{f}})|^{2}$ but for a cavity with higher effective reflectivity $|r_{m}| \approx 0.98$ ($N = 10$).}
\label{fig:spatial_photon}
\end{figure}

In Fig.~\ref{fig:photon_shape} we show the numerical solutions of dynamical equations~\eqref{eq:dyn_true} and~\eqref{eq:dyn_in_out}. We study an intermediate coupling regime, where the atom-cavity coupling is such that there are no Rabi oscillations between them and the photon leaks out of the cavity after being produced: $g_{m} \lesssim \textcolor{black}{\kappa}_{m}$. In the inside-outside representation the function $P_{f,1,0}\textcolor{black}{=|c_{f,1,0}(t)|^{2}}$ describes the shape of the produced photon inside the cavity and the function $P_{f,0,1}\textcolor{black}{=\int_{0}^{\infty} d\omega \, |c_{f,0,1}(\omega,t)|^{2}}$ corresponds to the photon outside the cavity, while in the true-mode representation, since there is no separation between the inside and the outside, there is one term describing the photon: $\tilde{P}_{f,1}\textcolor{black}{=\int_{0}^{\infty} d\omega \, |\tilde{c}_{f,1}(\omega,t)|^{2}}$, which incorporates the inside and the outside photon states, such that at the limit of validity of inside-outside representation $\tilde{P}_{f,1} \approx P_{f,1,0}+P_{f,0,1}$. In Fig.~\ref{fig:spectral_photon}, we compare the outgoing photon shape obtained via each representation in the frequency domain. {Additionally, taking into account the fact that one can obtain the spectral distribution of the photon via the Fourier transform of the state $c_{f,1,0}(t)$, we define a photon state}
\begin{eqnarray}
\color{black}
\hat{P}_\text{outs} = {\sigma} \Big \vert {\int^t_0 dt \, c^{\ast}_{f,1,0}(t) e^{-i\omega t}} \Big \vert^2,
\end{eqnarray}
{where $c_{f,1,0}(t)$ is the photon state inside the cavity in the inside-outside representation and $\sigma = \textcolor{black}{\kappa}_m/ 2\pi$ is the normalization constant, obtained by taking into account the fact that at $t \rightarrow \infty$ the photon is completely outside the cavity, such that $\int d\omega \hat{P}_\text{outs} = 1$.} As we can see from Fig.~\ref{fig:spectral_photon}, {at such a long time limit, in the case when the cavity mirror has enough layers such that we are in a regime of high effective reflectivity, the photon derived from the inside-outside representation and the corresponding Fourier transform definition} almost exactly matches with the photon shape derived from the true-mode representation. On the other hand, when the leakage through the multilayer mirror is strong, i.e., the effective reflectivity is relatively low, the mismatch between the two representations is significant. Additionally, in the regime where $\ell_c/\ell_0\neq \textcolor{black}{p}$, the mismatch is also significant. These mismatches are due to the fact that at these regimes the individual Lorentzians of the response function have overlaps and the emitter couples to more than one mode. In such an overlapped regime, the Lorentzians neighboring  the main peak can not be directly added to the inside-outside model as a separate mode $m'$, since, due to this overlap, the quantized modes associated to each of these Lorentzians will not satisfy the commutation relations. Nevertheless, this does not exclude a more general multimode inside--outside representation. One possible approach would be to construct the overlap matrix of the spectral functions associated with neighboring resonances and diagonalize it to obtain an orthonormal set of linear combinations. Canonical creation and annihilation operators could then be associated with these orthogonalized modes, which would generally no longer correspond to individual resonance peaks. The resulting effective Hamiltonian would require the inclusion of intermode couplings and, potentially, correlations induced by their interaction with a common external reservoir~\cite{Viviescas}. A complete derivation of such a multimode representation, including a consistent cavity--reservoir decomposition, is beyond the scope of the present work and will be investigated in a future study.

 Finally, in Fig.~\ref{fig:spatial_photon}, we study the photon shape along the propagation direction, corresponding to the term defined in Eq.~\eqref{eq:photon_space_coef}. From Fig.~\ref{fig:photon_shape} we can determine the time after which the photon is completely outside the cavity. Using the expression~\eqref{eq:photon_space_coef} with the solutions of Eq.~\eqref{eq:dyn_in_out}, we can describe the propagation of the photon outside the cavity. As we can see from Fig.~\ref{fig:spatial_photon}, for the given cavity length, the width of the spatial distribution of the single photon is of the order $\sim 200\ell_{c}$, and the better the effective reflectivity, the broader the distribution of the photon outside the cavity. Additionally, we highlight that the sharp start of the single photon distribution in space is due to the strict limitation imposed by the speed of light.

 \section{Conclusion}
 
We have extended the true-mode and inside--outside description developed in Ref.~\cite{Saharyan5.033056} to realistic cavities with multilayer dielectric output mirrors. Building on the multilayer cavity response and effective-length description established in Ref.~\cite{multilayer}, we have derived the corresponding coherent cavity--reservoir coupling function while retaining the mode dependence introduced by the longitudinal structure of the mirror. The resulting coupling contains the multilayer-dependent cavity linewidth and effective cavity length and therefore differs, in general, from the coupling obtained for an effectively zero-thickness mirror.  This captures the properties of the outside propagating photon generated in a realistic cavity with multilayered mirrors. This derivation is done in the limit where the structure of the cavity is such that the cavity resonances can be characterized via well-separated Lorentzian-like functions. In particular, we explicitly estimated the accuracy of such an approximation with respect to the cavity length and the number of dielectric layers forming the mirror.  We have shown that in this limit one can replace the multilayer mirror with an effective single-layered mirror of negligible thickness. We introduced the cavity response function for this replaced mirror, which, in general, is different from a single-layered cavity response function and incorporates the effects caused by the multilayer structure of the actual cavity. Based on this, we defined the effective reflectivity and showed that even in the regime where the effective reflectivity is high, i.e., the cavity is of high finesse, the cavity-reservoir coupling derived here is different from the one derived for a single-layered cavity. This is due to the fact that the multilayer cavity effective length $\ell_{\text{eff}}$ and the cavity geometric length $\ell_{c}$ are different; they coincide only when $\ell_{c}/\ell_{0} = \textcolor{black}{p}$, where $\textcolor{black}{p}$ is an integer.  The model derived here leads to a more general description of the cavity-reservoir system, where one can take into account the structure of the mirror and freely vary the frequency of light, the multilayer structure and the position of the cavity mirror. {While here we focus on studying the dynamics of \textcolor{black}{an atom coupled to} a  multilayer cavity, this analysis \textcolor{black}{equally applies to} other platforms featuring such multilayered structures, such as quantum dots in \textcolor{black}{micropillar} cavities~\cite{Giesz2016}}. Finally, we used this generalized model to study the spectral shape and the propagation of a single photon produced in a cavity. For applications to quantum networking, the multilayer corrections are particularly relevant when the effective cavity resonance frequencies differ appreciably from those predicted using the geometric cavity length. Such discrepancies can modify the detuning between an emitter and the dominant cavity resonance, affecting the spectrum of the generated photon and its subsequent matching to a remote quantum node. This is particularly useful when studying the reverse process of photon absorption and the full process of generation and absorption in more realistic cavity QED systems. Additionally, even though here we did not explicitly consider the laser field used to excite the atom, this model can be used to study the control of the photonic states produced from such cavities. 

{The use of quasi-normal modes \cite{Franke1,Franke2} as an hybrid basis to describe both the dynamics inside the cavity and the propagating photons outside offers interesting perspectives.}

%
%

\ack{We acknowledge H. R. Jauslin for helpful discussions. We acknowledge support from the European Union's Horizon 2020 research and innovation program under the Marie Sklodowska-Curie grant agreement No. 765075 (LIMQUET). A.S. and S.G. acknowledge additional support from EIPHI Graduate School (ANR-17-EURE-0002). A.S. acknowledges support from the postdoctoral fellowship of the IPD-STEMA program of the University of Liège.}

\bibliographystyle{iopart-num}

\bibliography{cav_photon}

\providecommand{\newblock}{}
\begin{thebibliography}{10}
\expandafter\ifx\csname url\endcsname\relax
  \def\url#1{{\tt #1}}\fi
\expandafter\ifx\csname urlprefix\endcsname\relax\def\urlprefix{URL }\fi
\providecommand{\eprint}[2][]{\url{#2}}

\bibitem{nielsen00}
Nielsen M~A and Chuang I~L 2000 {\em Quantum Computation and Quantum
  Information\/} (Cambridge University Press)

\bibitem{Kimble2008}
Kimble H~J 2008 {\em Nature\/} {\bf 453} 1023--1030 ISSN 1476-4687
  \urlprefix\url{https://doi.org/10.1038/nature07127}

\bibitem{cQED2}
Kuhn A, Hennrich M and Rempe G 2002 {\em Phys. Rev. Lett.\/} {\bf 89}(6) 067901
  \urlprefix\url{https://link.aps.org/doi/10.1103/PhysRevLett.89.067901}

\bibitem{ciracQuantumOpticsWhat2017a}
Cirac J~I and Kimble H~J 2017 {\em Nature Photonics\/} {\bf 11} 18--20 ISSN
  1749-4893

\bibitem{Cacciapuoti1}
Cacciapuoti A~S, Caleffi M, Tafuri F, Cataliotti F~S, Gherardini S and Bianchi
  G 2020 {\em IEEE Network\/} {\bf 34} 137--143

\bibitem{Ritter2012}
Ritter S, N{\"o}lleke C, Hahn C, Reiserer A, Neuzner A, Uphoff M, M{\"u}cke M,
  Figueroa E, Bochmann J and Rempe G 2012 {\em Nature\/} {\bf 484} 195--200
  ISSN 1476-4687 \urlprefix\url{https://doi.org/10.1038/nature11023}

\bibitem{Gorshkov}
Gorshkov A~V, Andr\'e A, Lukin M~D and S\o{}rensen A~S 2007 {\em Phys. Rev.
  A\/} {\bf 76}(3) 033804
  \urlprefix\url{https://link.aps.org/doi/10.1103/PhysRevA.76.033804}

\bibitem{McKeever}
McKeever J, Boca A, Boozer A~D, Miller R, Buck J~R, Kuzmich A and Kimble H~J
  2004 {\em Science\/} {\bf 303} 1992--1994
  \urlprefix\url{https://www.science.org/doi/abs/10.1126/science.1095232}

\bibitem{cQED4}
Kuhn A and Ljunggren D 2010 {\em Contemporary Physics\/} {\bf 51} 289--313

\bibitem{Jaksch2000}
Jaksch D, Cirac J~I, Zoller P, Rolston S~L, C\^ot\'e R and Lukin M~D 2000 {\em
  Phys. Rev. Lett.\/} {\bf 85}(10) 2208--2211
  \urlprefix\url{https://link.aps.org/doi/10.1103/PhysRevLett.85.2208}

\bibitem{Saffman2010}
Saffman M, Walker T~G and M\o{}lmer K 2010 {\em Rev. Mod. Phys.\/} {\bf 82}(3)
  2313--2363
  \urlprefix\url{https://link.aps.org/doi/10.1103/RevModPhys.82.2313}

\bibitem{Keller2004}
Keller M, Lange B, Hayasaka K, Lange W and Walther H 2004 {\em Nature\/} {\bf
  431} 1075--1078 ISSN 1476-4687
  \urlprefix\url{https://doi.org/10.1038/nature02961}

\bibitem{Stick2006}
Stick D, Hensinger W~K, Olmschenk S, Madsen M~J, Schwab K and Monroe C 2006
  {\em Nature Physics\/} {\bf 2} 36--39 ISSN 1745-2481
  \urlprefix\url{https://doi.org/10.1038/nphys171}

\bibitem{Ding}
Ding X, He Y, Duan Z~C, Gregersen N, Chen M~C, Unsleber S, Maier S, Schneider
  C, Kamp M, H\"ofling S, Lu C~Y and Pan J~W 2016 {\em Phys. Rev. Lett.\/} {\bf
  116}(2) 020401
  \urlprefix\url{https://link.aps.org/doi/10.1103/PhysRevLett.116.020401}

\bibitem{Yao}
Yao W, Liu R~B and Sham L~J 2005 {\em Phys. Rev. Lett.\/} {\bf 95}(3) 030504
  \urlprefix\url{https://link.aps.org/doi/10.1103/PhysRevLett.95.030504}

\bibitem{Somaschi2016}
Somaschi N, Giesz V, De~Santis L, Loredo J~C, Almeida M~P, Hornecker G,
  Portalupi S~L, Grange T, Ant{\'o}n C, Demory J, G{\'o}mez C, Sagnes I,
  Lanzillotti-Kimura N~D, Lema{\'i}tre A, Auffeves A, White A~G, Lanco L and
  Senellart P 2016 {\em Nature Photonics\/} {\bf 10} 340--345 ISSN 1749-4893
  \urlprefix\url{https://doi.org/10.1038/nphoton.2016.23}

\bibitem{moehringEntanglementSingleatomQuantum2007}
Moehring D~L, Maunz P, Olmschenk S, Younge K~C, Matsukevich D~N, Duan L~M and
  Monroe C 2007 {\em Nature\/} {\bf 449} 68--71 ISSN 1476-4687
  \urlprefix\url{https://doi.org/10.1038/nature06118}

\bibitem{barrettMultimodeInterferometryEntangling2019a}
Barrett T~D, Rubenok A, Stuart D, Barter O, Holleczek A, Dilley J,
  {Nisbet-Jones} P~B~R, Poulios K, Marshall G~D, O'Brien J~L, Politi A,
  Matthews J~C~F and Kuhn A 2019 {\em Quantum Science and Technology\/} {\bf 4}
  025008 ISSN 2058-9565

\bibitem{cQED1}
Ma\^{\i}tre X, Hagley E, Nogues G, Wunderlich C, Goy P, Brune M, Raimond J~M
  and Haroche S 1997 {\em Phys. Rev. Lett.\/} {\bf 79}(4) 769--772
  \urlprefix\url{https://link.aps.org/doi/10.1103/PhysRevLett.79.769}

\bibitem{Wilk}
Wilk T, Webster S~C, Kuhn A and Rempe G 2007 {\em Science (New York, N.Y.)\/}
  {\bf 317} 488—490 ISSN 0036-8075
  \urlprefix\url{https://doi.org/10.1126/science.1143835}

\bibitem{Boozer}
Boozer A~D, Boca A, Miller R, Northup T~E and Kimble H~J 2007 {\em Phys. Rev.
  Lett.\/} {\bf 98}(19) 193601
  \urlprefix\url{https://link.aps.org/doi/10.1103/PhysRevLett.98.193601}

\bibitem{Dilley}
Dilley J, Nisbet-Jones P, Shore B~W and Kuhn A 2012 {\em Phys. Rev. A\/} {\bf
  85}(2) 023834
  \urlprefix\url{https://link.aps.org/doi/10.1103/PhysRevA.85.023834}

\bibitem{Law}
Law C~K and Kimble H~J 1997 {\em Journal of Modern Optics\/} {\bf 44}
  2067--2074
  \urlprefix\url{https://www.tandfonline.com/doi/abs/10.1080/09500349708231869}

\bibitem{An_PhysRevA.79.032303}
An J~H, Feng M and Oh C~H 2009 {\em Phys. Rev. A\/} {\bf 79}(3) 032303
  \urlprefix\url{https://link.aps.org/doi/10.1103/PhysRevA.79.032303}

\bibitem{Barak1152261}
Dayan B, Parkins A~S, Aoki T, Ostby E~P, Vahala K~J and Kimble H~J 2008 {\em
  Science\/} {\bf 319} 1062--1065 (\textit{Preprint}
  \eprint{https://www.science.org/doi/pdf/10.1126/science.1152261})
  \urlprefix\url{https://www.science.org/doi/abs/10.1126/science.1152261}

\bibitem{Jaynes1963}
Jaynes E~T and Cummings F~W 1963 {\em Proceedings of the IEEE\/} {\bf 51}
  89--109 \urlprefix\url{https://ieee.org}

\bibitem{Eberly1980}
Eberly J~H, Narozhny N~B and Sanchez-Mondragon J~J 1980 {\em Phys. Rev.
  Lett.\/} {\bf 44}(20) 1323--1326
  \urlprefix\url{https://link.aps.org/doi/10.1103/PhysRevLett.44.1323}

\bibitem{BARNETT1988364}
Barnett S and Radmore P 1988 {\em Optics Communications\/} {\bf 68} 364--368
  ISSN 0030-4018
  \urlprefix\url{https://www.sciencedirect.com/science/article/pii/0030401888902337}

\bibitem{gardiner00}
Gardiner C~W and Zoller P 2004 {\em Quantum Noise\/} 3rd ed (Springer)

\bibitem{Breuer}
Breuer H~P and Petruccione F 2002 {\em The theory of open quantum systems\/}
  (Great Clarendon Street: Oxford University Press)

\bibitem{Dutra2000}
Dutra S~M and Nienhuis G 2000 {\em Journal of Optics B: Quantum and
  Semiclassical Optics\/} {\bf 2} 584
  \urlprefix\url{https://doi.org/10.1088/1464-4266/2/5/305}

\bibitem{dutra2005cavity}
Dutra S~M 2005 {\em Cavity quantum electrodynamics: the strange theory of light
  in a box\/} (Wiley, Hoboken, NJ)

\bibitem{pseudomodes}
Dalton B~J, Barnett S~M and Garraway B~M 2001 {\em Phys. Rev. A\/} {\bf 64}(5)
  053813 \urlprefix\url{https://link.aps.org/doi/10.1103/PhysRevA.64.053813}

\bibitem{Viviescas}
Viviescas C and Hackenbroich G 2003 {\em Phys. Rev. A\/} {\bf 67}(1) 013805
  \urlprefix\url{https://link.aps.org/doi/10.1103/PhysRevA.67.013805}

\bibitem{Lentrodt}
Lentrodt D and Evers J 2020 {\em Phys. Rev. X\/} {\bf 10}(1) 011008
  \urlprefix\url{https://link.aps.org/doi/10.1103/PhysRevX.10.011008}

\bibitem{Tamascelli}
Tamascelli D, Smirne A, Huelga S~F and Plenio M~B 2018 {\em Phys. Rev. Lett.\/}
  {\bf 120}(3) 030402
  \urlprefix\url{https://link.aps.org/doi/10.1103/PhysRevLett.120.030402}

\bibitem{Medina}
Medina I, Garc\'{\i}a-Vidal F~J, Fern\'andez-Dom\'{\i}nguez A~I and Feist J
  2021 {\em Phys. Rev. Lett.\/} {\bf 126}(9) 093601
  \urlprefix\url{https://link.aps.org/doi/10.1103/PhysRevLett.126.093601}

\bibitem{Pleasance}
Pleasance G, Garraway B~M and Petruccione F 2020 {\em Phys. Rev. Res.\/} {\bf
  2}(4) 043058
  \urlprefix\url{https://link.aps.org/doi/10.1103/PhysRevResearch.2.043058}

\bibitem{Semin}
Semin G, Jauslin H~R and Gu\'erin S 2025 {\em Phys. Rev. A\/} {\bf 112}(4)
  043501 \urlprefix\url{https://link.aps.org/doi/10.1103/73rb-yr5m}

\bibitem{Franke1}
Franke S, Hughes S, Dezfouli M~K, Kristensen P~T, Busch K, Knorr A and Richter
  M 2019 {\em Phys. Rev. Lett.\/} {\bf 122}(21) 213901
  \urlprefix\url{https://link.aps.org/doi/10.1103/PhysRevLett.122.213901}

\bibitem{Franke2}
Franke S, Richter M, Ren J, Knorr A and Hughes S 2020 {\em Phys. Rev. Res.\/}
  {\bf 2}(3) 033456
  \urlprefix\url{https://link.aps.org/doi/10.1103/PhysRevResearch.2.033456}

\bibitem{Saharyan5.033056}
Saharyan A, Rousseaux B, Kis Z, Stryzhenko S and Gu\'erin S 2023 {\em Phys.
  Rev. Res.\/} {\bf 5}(3) 033056
  \urlprefix\url{https://link.aps.org/doi/10.1103/PhysRevResearch.5.033056}

\bibitem{vogel2006quantum}
Vogel W and Welsch D~G 2006 {\em Quantum optics\/} 3rd ed (Wiley-VCH) ISBN
  978-3-527-40507-7

\bibitem{multilayer}
Saharyan A, Álvarez J~R, Doherty T~H, Kuhn A and Guérin S 2021 {\em Applied
  Physics Letters\/} {\bf 118} 154002
  \urlprefix\url{https://doi.org/10.1063/5.0047145}

\bibitem{Knoll}
Kn\"oll L, Vogel W and Welsch D~G 1991 {\em Phys. Rev. A\/} {\bf 43}(1)
  543--553 \urlprefix\url{https://link.aps.org/doi/10.1103/PhysRevA.43.543}

\bibitem{murray1978laser}
Murray I~S, Scully M~O and Willis~E J~L 1978 {\em Laser Physics\/} (Avalon
  Publishing) ISBN 9780201069037
  \urlprefix\url{https://books.google.fr/books?id=gpuqswEACAAJ}

\bibitem{Federico_2022}
Federico M, Dorier V, Guérin S and Jauslin H~R 2022 {\em Journal of Physics B:
  Atomic, Molecular and Optical Physics\/} {\bf 55} 174002
  \urlprefix\url{https://dx.doi.org/10.1088/1361-6455/ac7e0e}

\bibitem{garrison2008quantum}
Garrison J and Chiao R 2008 {\em Quantum Optics\/} Oxford Graduate Texts (OUP
  Oxford) ISBN 9780198508861
  \urlprefix\url{https://books.google.fr/books?id=RCoSDAAAQBAJ}

\bibitem{Mandel_1966}
Mandel L 1966 {\em Phys. Rev.\/} {\bf 144}(4) 1071--1077
  \urlprefix\url{https://link.aps.org/doi/10.1103/PhysRev.144.1071}

\bibitem{Federico_2023_2}
Federico M and Jauslin H~R 2023 {\em Phys. Rev. A\/} {\bf 108}(4) 043720
  \urlprefix\url{https://link.aps.org/doi/10.1103/PhysRevA.108.043720}

\bibitem{Giesz2016}
Giesz V, Somaschi N, Hornecker G, Grange T, Reznychenko B, De~Santis L, Demory
  J, Gomez C, Sagnes I, Lema{\^i}tre A, Krebs O, Lanzillotti-Kimura N~D, Lanco
  L, Auffeves A and Senellart P 2016 {\em Nature Communications\/} {\bf 7}
  11986 ISSN 2041-1723 \urlprefix\url{https://doi.org/10.1038/ncomms11986}

\bibitem{Tomas}
Toma\ifmmode~\check{s}\else \v{s}\fi{} M~S 1995 {\em Phys. Rev. A\/} {\bf
  51}(3) 2545--2559
  \urlprefix\url{https://link.aps.org/doi/10.1103/PhysRevA.51.2545}

\end{thebibliography}

  \appendix
 
 \section{Multilayer cavity response and its effective approximation \label{app:equivalcne_single_layer}}
 
As it is shown in Ref.~\cite{multilayer} {(see also \cite{Tomas})}, the response function of a cavity having a multilayer mirror can be written as follows:
\begin{eqnarray}
\mathcal{T}_{\omega} = \frac{e^{-i\frac{\omega}{c}(\ell_{c}+(N-1)(\delta+\alpha))}}{B^{\ast}_{2N-2}(\omega)}\frac{t_{1}(\omega)}{1+e^{i\frac{\omega}{c}\delta}e^{{i\phi_{B}(\omega)}}r_{1}(\omega)}, 
\label{eq:resp_actual}
\end{eqnarray}
where $B_{2N-2}(\omega)$ is a term describing the multilayer nature of the cavity and is calculated {via the recursive boundary matching of the field at the dielectric interfaces}, $\phi_{B}(\omega) =  B_{2N-2}/B^{\ast}_{2N-2} $, the parameters $\delta$ and $\alpha$ are the lengths corresponding to each layer of the dielectric layer pair, and  $t_{1}(\omega)$ and $r_{1}(\omega)$ are the spectral transmission and reflection functions of a single layer having a thickness $\delta$  (assuming that the layer having a thickness $\alpha$ is vacuum) ~\cite{vogel2006quantum,multilayer}.

{The same response can be expressed in terms of the complex reflection and transmission amplitudes of the entire multilayer stack, as in the conventional transfer-matrix formulation. Indeed, the recurrence for $B_{j}(\omega,x)$ involves linear combinations of $B_{j-1}$ and $B_{j-1}^{\ast}$~\cite{multilayer}. Since $B_0(\omega,x)=e^{i\omega(x+\ell_c)/c}$, it follows that
\begin{align*}
B_{j}(\omega,x)&=u_{j}(\omega,x)e^{i\frac{\omega}{c}\ell_{c}}+v_{j}(\omega,x)e^{-i\frac{\omega}{c}\ell_{c}}
\end{align*}
where $u_j$ and $v_j$ are independent of $\ell_c$ and are obtained from the same boundary-matching recurrence~\cite{multilayer}. In particular, for the term $B_{2N-2}(\omega)$ we have 
\begin{align*}
B_{2N-2}(\omega) = u_{2N-2}\bigl(\omega,(N-1)(\delta+\alpha)\bigr)e^{i\frac{\omega}{c}\ell_{c}}+v_{2N-2}\bigl(\omega,(N-1)(\delta+\alpha)\bigr)e^{-i\frac{\omega}{c}\ell_{c}}.
\end{align*}
Substituting this decomposition into Eq.~\eqref{eq:resp_actual} gives
\begin{align*}
\mathcal{T}_{\omega} = \frac{t_{1}(\omega)e^{-i\frac{\omega}{c}((N-1)(\delta+\alpha))}}{D_{2N-2}(\omega)+C_{2N-2}(\omega)e^{2i\frac{\omega}{c}\ell_{c}}},
\end{align*}
where 
\begin{align*}
D_{2N-2}(\omega)&=u^{\ast}_{2N-2}(\omega,(N-1)(\delta+\alpha))+v_{2N-2}(\omega,(N-1)(\delta+\alpha))e^{i\frac{\omega}{c}\delta}r_{1}(\omega)\\
C_{2N-2}(\omega)&=v^{\ast}_{2N-2}(\omega,(N-1)(\delta+\alpha))+u_{2N-2}(\omega,(N-1)(\delta+\alpha))e^{i\frac{\omega}{c}\delta}r_{1}(\omega).
\end{align*}
Introducing the physical stack thickness $d_{\mathrm{stack}}=N\delta+(N-1)\alpha$, we define
\begin{align*}
r_{\textrm{multi}}(\omega)&=e^{-i\frac{\omega}{c} d_{\mathrm{stack}}}\frac{C_{2N-2}(\omega)}{D_{2N-2}(\omega)}\\
t_{\textrm{multi}}(\omega) &= e^{-i\frac{\omega}{c} d_{\mathrm{stack}}}e^{i\frac{\omega}{c}\delta}\frac{t_{1}(\omega)}{D_{2N-2}(\omega)}.
\end{align*}
The reflection phase is thus referred to the midpoint of the stack, in analogy with the single-layer convention of Ref.~\cite{multilayer}. The exact cavity response becomes
\begin{align}
\mathcal{T}_{\omega} = \frac{t_{\textrm{multi}}(\omega)}{1+{r}_{\textrm{multi}}(\omega)e^{2i\frac{\omega}{c}\left(\ell_{c}+\frac{d_{\mathrm{stack}}}{2}\right)}} =\mathcal{T}_{\omega} = \frac{t_{\textrm{multi}}(\omega)}{1+{r}_{\textrm{multi}}(\omega)e^{2i\frac{\omega}{c}\ell_{\textrm{ref}}}},
\label{eq:response_with_multi_transm_reflect}
\end{align}
where we have defined $\ell_{\textrm{ref}} = \ell_{c}+\frac{d_{\mathrm{stack}}}{2}$. With this, the round-trip phase can be written as $\Phi_{\textrm{round}}(\omega) = 2\frac{\omega}{c}\ell_{\textrm{ref}}+\varphi_{\textrm{multi}}(\omega)$, where $\varphi_{\textrm{multi}}(\omega) = \textrm{arg}\left\{r_{\textrm{multi}}(\omega)\right\}$. We note that for the present lossless, symmetric stack, these coefficients obey
\begin{align*}
|t_{\textrm{multi}}(\omega)|^{2}+|r_{\textrm{multi}}(\omega)|^{2} = 1,\\
r^{*}_{\mathrm{multi}}t_{\mathrm{multi}}+t^{*}_{\mathrm{multi}}r_{\mathrm{multi}}=0.
\end{align*}
Hence the $B_j$ recurrence yields the same complex scattering response as the transfer-matrix approach, while also providing the spatial true modes required for the subsequent field quantization. \\
We next relate this physical multilayer response to the parameters of the Lorentzian approximation employed in the main text. Following the expansion of single-layered mirror response function demonstrated in Refs.~\cite{dutra2005cavity,multilayer}, we can write the following: 
\begin{align}
\nonumber
|\mathcal{T}_{\omega}|^{2}&=\frac{\left|t_{\textrm{multi}}(\omega)\right|^{2}}{\left|1+|r_{\textrm{multi}}(\omega)|e^{i\left(2\frac{\omega}{c}\ell_{\textrm{ref}}+\varphi_{\textrm{multi}}(\omega)\right)}\right|^{2}}=\frac{1-\left|r_{\textrm{multi}}(\omega)\right|^{2}}{\left|1+|r_{\textrm{multi}}(\omega)|e^{i\left(2\frac{\omega}{c}\ell_{\textrm{ref}}+\varphi_{\textrm{multi}}(\omega)\right)}\right|^{2}}\\ \nonumber
&=\left[1-\frac{\left|r_{\textrm{multi}}(\omega)\right|e^{i\left(2\frac{\omega}{c}\ell_{\textrm{ref}}+\varphi_{\textrm{multi}}(\omega)\right)}}{1+|r_{\textrm{multi}}(\omega)|e^{i\left(2\frac{\omega}{c}\ell_{\textrm{ref}}+\varphi_{\textrm{multi}}(\omega)\right)}}-\frac{\left|r_{\textrm{multi}}(\omega)\right|e^{-i\left(2\frac{\omega}{c}\ell_{\textrm{ref}}+\varphi_{\textrm{multi}}(\omega)\right)}}{1+|r_{\textrm{multi}}(\omega)|e^{-i\left(2\frac{\omega}{c}\ell_{\textrm{ref}}+\varphi_{\textrm{multi}}(\omega)\right)}}\right]\\ \nonumber
&=\left[1+\sum^{\infty}_{k=1}\left|r_{\textrm{multi}}(\omega)\right|^{k}\left(e^{ik\left(2\frac{\omega}{c}\ell_{\textrm{ref}}+\varphi_{\textrm{multi}}(\omega)-\pi\right)}+e^{-ik\left(2\frac{\omega}{c}\ell_{\textrm{ref}}+\varphi_{\textrm{multi}}(\omega)-\pi\right)}\right)\right]\\ \nonumber
&=\sum^{\infty}_{k=-\infty}\left|r_{\textrm{multi}}(\omega)\right|^{|k|}e^{ik\left(2\frac{\omega}{c}\ell_{\textrm{ref}}+\varphi_{\textrm{multi}}(\omega)-\pi\right)}\\ \nonumber
&=\sum^{\infty}_{k=-\infty}\frac{-2\ln\left|r_{\textrm{multi}}(\omega)\right|}{\left(\ln\left|r_{\textrm{multi}}(\omega)\right|\right)^{2}+\left(2\frac{\omega}{c}\ell_{\textrm{ref}}+\varphi_{\textrm{multi}}(\omega)-\pi-2\pi k\right)^{2}}\\ \nonumber
&=\sum^{\infty}_{k=-\infty}\left(\frac{c}{2\ell_{\textrm{ref}}}\right)^{2}\frac{-2\ln\left|r_{\textrm{multi}}(\omega)\right|}{\left(\ln\left|r_{\textrm{multi}}(\omega)\right|\frac{c}{2\ell_{\textrm{ref}}}\right)^{2}+\left(\omega+\frac{c}{2\ell_{\textrm{ref}}}\left(\varphi_{\textrm{multi}}(\omega)-\pi-2\pi k\right)\right)^{2}}\\ \label{eq:multilayer_poisson}
&=\sum^{\infty}_{k=-\infty}\frac{c}{2\ell_{\textrm{ref}}}\frac{\gamma_{\textrm{multi}}(\omega)}{\left(\frac{\gamma_{\textrm{multi}}(\omega)}{2}\right)^{2}+\left(\omega-\tilde{\omega}_{k}(\omega)\right)^{2}},
\end{align}
where the Poisson-summation identity has been used and
\begin{align*}
\gamma_{\textrm{multi}}(\omega)&=-\frac{c}{\ell_{\textrm{ref}}}\ln\left|r_{\textrm{multi}}(\omega)\right|\\
\tilde{\omega}_{k}(\omega)&=\frac{c\pi k}{\ell_{\textrm{ref}}}-\frac{c}{2\ell_{\textrm{ref}}}\left(\varphi_{\textrm{multi}}(\omega)-\pi\right).
\end{align*}
Eq.~\eqref{eq:multilayer_poisson} is an exact identity, its individual terms are not yet Lorentzians with frequency-independent centers and linewidths. \\
\begin{figure}[!h]
\centering
\includegraphics[width=0.92\linewidth]{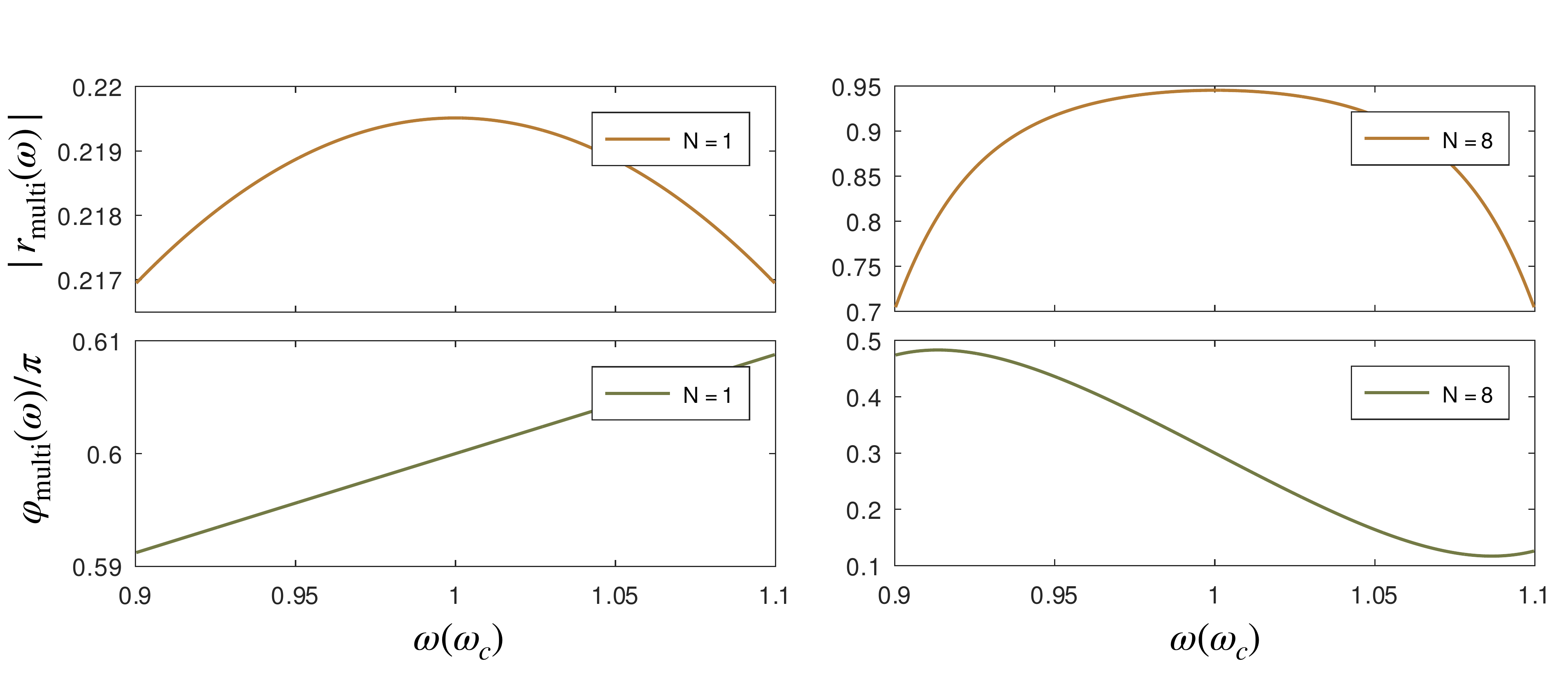}
\caption{Magnitude (upper panels) and phase (lower panels) of the reflection amplitude $r_{\mathrm{multi}}(\omega)$ of the physical dielectric stack for $N=1$ (left) and $N=8$ (right). The frequency is normalized by the design frequency $\omega_c=2\pi c/\lambda_0$; the phase is shown in units of $\pi$. The curves correspond to the quarter-wave configuration $\delta=\lambda_0/(4n_1)$ and $\alpha=\lambda_0/4$ with $n_1=1.25$~\cite{multilayer}. Note that the vertical scales differ between the two columns. The reflection phase alone need not equal $\pi$ at a cavity resonance; it is the total round-trip phase that enters the resonance condition.}
\label{fig:phase_plot}
\end{figure}
Figure~\ref{fig:phase_plot} illustrates the frequency dependence of the physical multilayer reflection amplitude. In particular, the reflection magnitude is smooth in the vicinity of the design frequency, although it can vary appreciably across the wider spectral interval, especially with increasing $N$. For a sufficiently narrow and isolated cavity resonance (see for example Fig.~\ref{fig:fig1} of the main text), its variation over the corresponding linewidth may therefore be neglected, $|r_{\textrm{multi}}(\omega)|\simeq \left|r_{\textrm{multi}}(\omega_N^{(m)})\right|$. At this narrowband regime $\omega_N^{(m)}$, which here is obtained numerically, coincides with the frequency corresponding to the maximum of $|\mathcal{T}_{\omega}|^{2}$  over the region around each peak. This corresponds to the condition that the round-trip phase $\Phi_{\mathrm{round}}(\omega_N^{(m)}) = \pi+2\pi k_m$, making $\tilde{\omega}_{k_m}(\omega_N^{(m)})=\omega_N^{(m)}$. Expanding the frequency-dependent term in Eq.~\eqref{eq:multilayer_poisson} to first order about $\omega_{N}^{(m)}$ yields
\begin{align*}
\omega-\tilde{\omega}_{k_m}(\omega) &= \left[1-\left.\frac{d\tilde{\omega}_{k}(\omega)}{d\omega} \right|_{\omega_N^{(m)}}\right](\omega-\omega_N^{(m)}) = \left[1+\frac{c}{2\ell_{\textrm{ref}}}\left.\frac{d\varphi_{\textrm{multi}}(\omega)}{d\omega} \right|_{\omega_N^{(m)}}\right](\omega-\omega_N^{(m)})\\
&= \frac{L_{N}^{(m)} }{\ell_{\textrm{ref}} }(\omega-\omega_N^{(m)}),
\end{align*}
where 
\begin{align*}
L_{N}^{(m)} =\ell_{\textrm{ref}} +\frac{c}{2}\left.\frac{d\varphi_{\textrm{multi}}(\omega)}{d\omega} \right|_{\omega_N^{(m)}}
\end{align*} 
characterizes the first-order frequency dependence of the total round-trip phase. Under these conditions the local contribution of an isolated resonance becomes
\begin{align}
|\mathcal{T}_{\omega}|^{2}&\simeq  \frac{c}{2L^{(m)}_{N}}\frac{{\kappa}_N^{(m)}}{(\omega-\omega^{(m)}_{N})^{2}+[{\kappa}_N^{(m)}/2]^{2}},\qquad{\kappa}_N^{(m)}=-\frac{c}{L^{(m)}_{N}}\ln|r_{\mathrm{multi}}(\omega^{(m)}_{N})|.
\label{eq:multilayer_local_fit}
\end{align}
This demonstrates that the Lorentzian parameters depend not only on the magnitude of the physical multilayer reflectivity, but also on the first derivative of its phase. Higher-order phase dispersion can be retained systematically, but generally produces deviations from the symmetric Lorentzian profile that cannot be absorbed into a constant $L_N^{(m)}$.\\
We now relate this local description to the fictitious zero-thickness mirror used in the effective inside--outside model. For a selected fitted resonance, we define $\ell^{(m)}_{\mathrm{eff}}=m\pi c/\omega^{(m)}_{N}$. The effective round-trip phase is $\Phi_{\mathrm{eff}}(\omega)=2\omega\ell_{\mathrm{eff}}^{(m)}/c+\varphi_m(\omega)$. We choose $\varphi_m(\omega_N^{(m)})=\pi$ (modulo $2\pi$), so that the selected resonance satisfies the phase condition in this effective
cavity. Thus, once the physical cavity, the selected resonance, and its mode index have been specified, the effective mirror position is fixed by construction. The corresponding effective reflection magnitude is defined from the linewidth as $|r_{m}(\omega^{(m)}_{N})|=\exp\!\left[-\frac{\kappa^{(m)}_{N}\ell^{(m)}_{\mathrm{eff}}}{c}\right]$. Using Eq.~\eqref{eq:multilayer_local_fit}, its relation to the physical multilayer reflectivity is
\begin{align*}
|r_{m}(\omega^{(m)}_{N})|=|r_{\mathrm{multi}}(\omega^{(m)}_{N})|^{\frac{\ell^{(m)}_{\mathrm{eff}}}{L^{(m)}_{N}}}.
\end{align*}
The phase choice of the fictitious mirror is consistent with that of the physical multilayer after referring the latter to the effective plane: the corresponding phase is
\begin{align*}
\varphi_{m}(\omega)=\varphi_{\mathrm{multi}}(\omega)+\frac{2\omega}{c}\bigl(\ell_{\mathrm{ref}}-\ell^{(m)}_{\mathrm{eff}}\bigr),
\end{align*}
which equals $\pi$ at the resonance $\omega^{(m)}_{N}$. This equality, however, holds true only around the selected resonance. Additionally, for such a symmetric, lossless fictitious mirror we impose $|t_m(\omega)|^2+|r_m(\omega)|^2=1$ and $r_m^*(\omega)t_m(\omega)+t_m^*(\omega)r_m(\omega)=0$. With these the local effective response can be written as
\begin{align}
\label{eq:app:resp_app_single}
{T}_{m,\omega} = \sqrt{\frac{\ell^{(m)}_{\text{eff}}}{L^{(m)}_{N}}}\frac{t_{m}(\omega)}{1+r_{m}(\omega)e^{2i\frac{\omega}{c}\ell^{(m)}_{\text{eff}}}}. 
\end{align}
If $\kappa_N^{(m)}\ell_{\mathrm{eff}}^{(m)}/c\ll1$, (which corresponds to well separated Lorentzian peaks) and the effective reflection amplitude varies negligibly across the selected linewidth, its spectral intensity takes the Lorentzian form derived above 
\begin{align*}
 |{T}_{m,\omega}|^2\simeq {\frac{c}{2L^{(m)}_{N}}}\frac{{{\kappa}^{(m)}_{N}}}{\left(\omega-\omega^{(m)}_{N}\right)^2+\left(\frac{{\kappa}^{(m)}_{N}}{2}\right)^2} ,
\end{align*}
which at this regime can be further approximated as 
\begin{align*}
{T}_{m,\omega} \approx \sqrt{\frac{c}{2L^{(m)}_{N}}}\frac{\sqrt{{\kappa}^{(m)}_{N}}}{\left(\omega-\omega^{(m)}_{N}\right)+i\frac{{\kappa}^{(m)}_{N}}{2}}
\end{align*}
The above construction fixes the resonant effective reflectivity and reproduces the local spectral intensity, but it does not uniquely determine the off-resonant complex functions $r_m(\omega)$ and $t_m(\omega)$. }

\section{Discrete cavity modes in terms of to the true modes\label{app:in_operator}}

In this appendix, we derive the relation between the cavity operator in the inside-outside representation $a_{m}$ and the true modes $a_{\omega}$. From the equations 
\begin{align*}
\mathbb{E}_{\text{ins}}(x)&={E}_{\text{ins}}(x),\\
\mathbb{B}_{\text{ins}}(x)&={B}_{\text{ins}}(x)
\end{align*}
 we have 
 \begin{subequations}
 \begin{align}
\nonumber
\sum_{m=1}^{\infty}   \sqrt{\frac{\hbar\omega_m}{L_{N}\mathcal{A}\epsilon_0}}\left({a}_{m} + {a}^{\dag}_{m}\right)\sin{\left [\frac{\omega_m}{c}(x+\ell_c)\right]} =& \int_0^{\infty} d\omega \sqrt{\frac{\hbar\omega}{\pi c\mathcal{A}\epsilon_0}}\Big( \left[\sin\left[{\textstyle\frac{\omega}{c}}(x+\ell_c)\right]e^{i\frac{\omega}{c}\ell_c}\mathcal{T}_{\omega}\right] {a}_{\omega} \\
&+\left[\sin\left[{\textstyle\frac{\omega}{c}}(x+\ell_c)\right]e^{-i\frac{\omega}{c}\ell_c}\mathcal{T}^{\ast}_{\omega}\right] {a}_{\omega}^{\dagger}\Big), \\ 
\nonumber
\frac{i}{c}\sum_{m=1}^{\infty}    \sqrt{\frac{\hbar\omega_m}{L_{N}\mathcal{A}\epsilon_0}}\left({a}_{m} - {a}^{\dag}_{m}\right)\cos{\left [\frac{\omega_m}{c}(x+\ell_c)\right]}=&\,\frac{i}{c}\int_0^{\infty} d\omega \sqrt{\frac{\hbar\omega}{\pi c\mathcal{A}\epsilon_0}}\Big( \cos\left[{\textstyle\frac{\omega}{c}}(x+\ell_c)\right] e^{i\frac{\omega}{c}\ell_c}\mathcal{T}_{\omega}{a}_{\omega}\\
&  - \cos\left[{\textstyle\frac{\omega}{c}}(x+\ell_{c})\right]e^{-i\frac{\omega}{c}\ell_c}\mathcal{T}^{\ast}_{\omega} {a}^{\dag}_{\omega} \Big). 
\end{align}
\end{subequations}
which simplifies to 
\begin{subequations}
 \begin{align}
 \label{eq:equation_el}
\sum_{m=1}^{\infty}   \sqrt{\frac{\omega_m}{L_{N}}}({a}_{m} + {a}^{\dag}_{m})\sin{\left [\frac{\omega_m}{c}(x+\ell_c)\right]} =& \int_0^{\infty} d\omega \sqrt{\frac{\omega}{\pi c}}\Big( \left[\sin[{\textstyle\frac{\omega}{c}}(x+\ell_c)]e^{i\frac{\omega}{c}\ell_c}\mathcal{T}_{\omega}\right] {a}_{\omega} \\ \nonumber
&+ \left[\sin[{\textstyle\frac{\omega}{c}}(x+\ell_c)]e^{-i\frac{\omega}{c}\ell_c}\mathcal{T}^{\ast}_{\omega}\right] {a}_{\omega}^{\dagger}\Big), \\ 
\label{eq:equation_mag}
\sum_{m=1}^{\infty}    \sqrt{\frac{\omega_m}{L_{N}}}\left({a}_{m} - {a}^{\dag}_{m}\right)\cos{\left [\frac{\omega_m}{c}(x+\ell_c)\right]}=&\int_0^{\infty} d\omega \sqrt{\frac{\omega}{\pi c}}\Big( \cos\left[{\textstyle\frac{\omega}{c}}(x+\ell_c)\right] e^{i\frac{\omega}{c}\ell_c}\mathcal{T}_{\omega}{a}_{\omega} \\ \nonumber
&-\cos\left[{\textstyle\frac{\omega}{c}}(x+\ell_{c})\right]e^{-i\frac{\omega}{c}\ell_c}\mathcal{T}^{\ast}_{\omega} {a}^{\dag}_{\omega} \Big). 
\end{align}
\end{subequations}
 If we multiply both sides of eq.~\eqref{eq:equation_el} by $\sin{\left [\frac{\omega_n}{c}(x+\ell_c)\right]}$ and integrate over the space $x=(-\ell_{c},d)$, where $d =\ell_{\text{eff}}-\ell_{c}$, we get
 \begin{align}
 \label{eq:app:a_plus}
 \left({a}_{m} + {a}^{\dag}_{m}\right) =&\frac{1}{\ell_{\text{eff}}} \sqrt{\frac{c L_{N}}{\pi }}\int_0^{\infty} d\omega \sqrt{\frac{\omega}{\omega_m}}\left(\frac{\sin{\left[( \omega-\omega_m)\frac{\ell_{\text{eff}}}{c}\right]}}{\omega-\omega_m}-\frac{\sin{\left[(\omega+\omega_m)\frac{\ell_{\text{eff}}}{c}\right]}}{\omega+\omega_m}
 \right)\\ \nonumber
 &\times \left( e^{i\frac{\omega}{c}\ell_c}\mathcal{T}_{\omega} {a}_{\omega} +e^{-i\frac{\omega}{c}\ell_c}\mathcal{T}^{\ast}_{\omega} {a}_{\omega}^{\dagger}\right).
\end{align}
Similarly, if we multiply both sides of eq.~\eqref{eq:equation_mag} by $\cos{\left [\frac{\omega_n}{c}(x+\ell_c)\right]}$ and integrate we get
\begin{align}
\label{eq:app:a_minus}
 \left({a}_{m} - {a}^{\dag}_{m}\right) =&\frac{1}{\ell_{\text{eff}}} \sqrt{\frac{c L_{N}}{\pi }}\int_0^{\infty} d\omega \sqrt{\frac{\omega}{\omega_m}}\left(\frac{\sin{\left[( \omega-\omega_m)\frac{\ell_{\text{eff}}}{c}\right]}}{\omega-\omega_m}+\frac{\sin{\left[(\omega+\omega_m)\frac{\ell_{\text{eff}}}{c}\right]}}{\omega+\omega_m}
 \right)\\ \nonumber
 &\times \left( e^{i\frac{\omega}{c}\ell_c}\mathcal{T}_{\omega} {a}_{\omega} -e^{-i\frac{\omega}{c}\ell_c}\mathcal{T}^{\ast}_{\omega} {a}_{\omega}^{\dagger}\right).\end{align}
 By taking the sum of equations~\eqref{eq:app:a_plus} and~\eqref{eq:app:a_minus}, we get the expression for the discrete cavity operator ${a}_{m}$:
 \begin{align}
 {a}_{m}&=\frac{1}{\ell_{\text{eff}}}\sqrt{\frac{cL_{N}}{\pi}}\int d\omega\;\sqrt{\frac{\omega}{\omega_{m}}}\left(\frac{\sin\left[\left(\omega-\omega_{m}\right)\frac{\ell_{\text{eff}}}{c}\right]}{\omega-\omega_{m}}e^{i\frac{\omega}{c}\ell_{c}}\mathcal{T}_{\omega}{a}_{\omega}-\frac{\sin\left[\left(\omega+\omega_{m}\right)\frac{\ell_{\text{eff}}}{c}\right]}{\omega+\omega_{m}}e^{-i\frac{\omega}{c}\ell_{c}}\mathcal{T^{\ast}}_{\omega}{a}_{\omega}^{\dagger}\right).
 \end{align}

 \section{Continuous reservoir modes in terms of the true modes\label{app:out_operator}}
 
 Analogously to Appendix~\ref{app:in_operator}, in this one we derive the relation between the reservoir operator $b_{\omega}$ and the operator $a_{\omega}$. From the equations 
\begin{align*}
\mathbb{E}_{\text{outs}}(x)&={E}_{\text{outs}}(x),\\
\mathbb{B}_{\text{outs}}(x)&={B}_{\text{outs}}(x)
\end{align*}
 we get
 \begin{align}
\label{eq:electric_outside_equal}
2\int_0^{\infty} d\omega \sqrt{{\omega}}\Big({b}_{\omega} + {b}^{\dag}_{\omega}\Big)\sin{\Big [\frac{\omega}{c}(x-d)\Big]} &= -i\int_0^{\infty} d\omega \sqrt{{\omega}}\Big( \Big [e^{2i\frac{\omega}{c}\ell_c}\dfrac{\mathcal{T}_{\omega}}{\mathcal{T}^*_{\omega}}e^{i\frac{\omega}{c}x}-e^{-i\frac{\omega}{c}x} \Big ] {a}_{\omega} - H.c. \Big),\\ 
\label{eq:magnetic_outside_equal}
2\int_0^{\infty} d\omega \sqrt{{\omega}}\Big({b}_{\omega} - {b}^{\dag}_{\omega}\Big)\cos{\Big [\frac{\omega}{c}(x-d)\Big]}&=\int_0^{\infty} d\omega \sqrt{{\omega}}\Big( \Big [e^{2i\frac{\omega}{c}\ell_c}\dfrac{\mathcal{T}_{\omega}}{\mathcal{T}^*_{\omega}}e^{i\frac{\omega}{c}x}+e^{-i\frac{\omega}{c}x} \Big ] {a}_{\omega} - H.c. \Big)
\end{align}
We multiply both sides of Eq.~\eqref{eq:electric_outside_equal} by $\sin{\Big [\frac{\omega'}{c}(x-d)\Big]}$ and integrate over the space $(d,\infty)$, which gives:
\begin{align}
\Big({b}_{\omega'} + {b}^{\dagger}_{\omega'}\Big)&=-\frac{i}{{\pi c}}\int_d^{\infty} dx\int_0^{\infty} d\omega \sqrt{\frac{\omega}{\omega'}}\Big( \Big [e^{2i\frac{\omega}{c}\ell}\dfrac{\mathcal{T}_{\omega}}{\mathcal{T}^*_{\omega}}e^{i\frac{\omega}{c}x}-e^{-i\frac{\omega}{c}x} \Big ] {a}_{\omega}- H.c. \Big)\sin{\Big [\frac{\omega'}{c}(x-d)\Big]}.
\label{eq:sum_of_b}
\end{align}
Similarly, if we multiply both sides of equation~\eqref{eq:magnetic_outside_equal} by $\cos{\left[\frac{\omega'}{c}(x-d) \right]}$ we obtain:
\begin{align}
\label{eq:dif_of_b}
\left({b}_{\omega'} - {b}^{\dagger}_{\omega'}\right)&=\frac{1}{{\pi c}}\int_d^{\infty} dx\int_0^{\infty} d\omega \sqrt{\frac{\omega}{\omega'}}\Big( \Big [e^{2i\frac{\omega}{c}\ell_c}\dfrac{\mathcal{T}_{\omega}}{\mathcal{T}^*_{\omega}}e^{i\frac{\omega}{c}x}+e^{-i\frac{\omega}{c}x} \Big ] {a}_{\omega} - H.c. \Big)\cos{\left[\frac{\omega'}{c}(x-d) \right]}
\end{align}
If now we sum the expressions in~\eqref{eq:sum_of_b} and \eqref{eq:dif_of_b} we obtain:
\begin{equation}
\begin{split}
{b}_{\omega'}=&\,\frac{1}{2{\pi c}}\int_d^{\infty} dx\int_0^{\infty} d\omega \sqrt{\frac{\omega}{\omega'}} \left( \left [e^{2i\frac{\omega}{c}\ell_c}\dfrac{\mathcal{T}_{\omega}}{\mathcal{T}^*_{\omega}}e^{i\frac{\omega}{c}d}e^{i\frac{\omega-\omega'}{c}\left( x-d \right)}+e^{-i\frac{\omega}{c}d}e^{-i\frac{\omega-\omega'}{c}\left( x-d \right)} \right ] {a}_{\omega} \right.\\
&\left.- \left [e^{-2i\frac{\omega}{c}\ell_c}\dfrac{\mathcal{T}^{\ast}_{\omega}}{\mathcal{T}_{\omega}}e^{-i\frac{\omega}{c}d}e^{-i\frac{\omega+\omega'}{c}\left( x-d \right)}+e^{i\frac{\omega}{c}d}e^{i\frac{\omega+\omega'}{c}\left( x-d \right)} \right ] {a}^{\dagger}_{\omega} \right).
\end{split}
\end{equation}
 We define 
 \begin{align}
 G_{\omega}=e^{2i\frac{\omega}{c}\ell_{c}}\dfrac{\mathcal{T}_{\omega}}{\mathcal{T}_{\omega}^{\ast}}e^{i\frac{\omega}{c}d}.
 \end{align}
Considering the limit where $\mathcal{T}_{\omega} \approx \mathcal{T}_{m}(\omega)\approx T_{m,\omega}$, the above expression can be simplified using the formulation in Eq.~\eqref{eq:resp_sing_for_mult}:
 \begin{align*}
G_{\omega}	&\approx  e^{2i\frac{\omega}{c}\ell_{c}}\frac{{T}_{m,\omega}}{{T}_{m,\omega}^{\ast}}e^{i\frac{\omega}{c}d}\\
&=e^{2i\frac{\omega}{c}\ell_{c}}e^{i\frac{\omega}{c}d}\frac{t_{m}(\omega)}{t_{m}^{\ast}(\omega)}\frac{1+r_{m}^{\ast}(\omega)e^{-2i\frac{\omega}{c}\ell_{\text{eff}}}}{1+r_{m}(\omega)e^{2i\frac{\omega}{c}\ell_{\text{eff}}}}=- e^{2i\frac{\omega}{c}\ell_{c}}e^{i\frac{\omega}{c}d}e^{2i\varphi_{r}(\omega)}\frac{1+|r_{m}(\omega)|e^{-i\varphi_{r}(\omega)}e^{-2i\frac{\omega}{c}\ell_{\text{eff}}}}{1+|r_{m}(\omega)|e^{i\varphi_{r}(\omega)}e^{2i\frac{\omega}{c}\ell_{\text{eff}}}}\\[10pt]
	&=-e^{-i\frac{\omega}{c}d}\left(e^{2i\frac{\omega}{c}\ell_{c}}e^{2i\frac{\omega}{c}d}e^{2i\varphi_{r}(\omega)}\frac{1+|r_{m}(\omega)|e^{-i\left(\varphi_{r}(\omega)+2\frac{\omega}{c}\ell_{\text{eff}}\right)}}{1+|r_{m}(\omega)|e^{i\left(\varphi_{r}(\omega)+2\frac{\omega}{c}\ell_{\text{eff}}\right)}}+1-1\right)\\[10pt]
	&=	e^{-i\frac{\omega}{c}d}-e^{i\frac{\omega}{c}\ell_{c}}e^{i\varphi_{r}(\omega)}\frac{2\cos\left(\frac{\omega}{c}d+\frac{\omega}{c}\ell_{c}+\varphi_{r}(\omega)\right)+|r_{m}(\omega)|2\cos\left(-2\frac{\omega}{c}\ell_{\text{eff}}+\frac{\omega}{c}d+\frac{\omega}{c}\ell_{c}\right)}{1+|r_{m}(\omega)|e^{i\varphi_{r}(\omega)}e^{2i\frac{\omega}{c}\ell_{\text{eff}}}}\\[10pt]
	&=	e^{-i\frac{\omega}{c}d}+ i  e^{i\frac{\omega}{c}\ell_{c}}\frac{2\cos\left(\frac{\omega}{c}d+\frac{\omega}{c}\ell_{c}+\varphi_{r}(\omega)\right)+|r_{m}(\omega)|2\cos\left(-2\frac{\omega}{c}\ell_{\text{eff}}+\frac{\omega}{c}d+\frac{\omega}{c}\ell_{c}\right)}{|t_{m}(\omega)|}{T}_{m,\omega}\sqrt{\frac{L_{N}}{\ell_{\text{eff}}}}\\[10pt]
	&=e^{-i\frac{\omega}{c}d}+ 2ie^{i\frac{\omega}{c}\ell_{c}}\frac{\cos\left(\frac{\omega}{c}\ell_{\text{eff}}+\varphi_{r}(\omega)\right)+|r_{m}(\omega)|\cos\left(\frac{\omega}{c}\ell_{\text{eff}}\right)}{|t_{m}(\omega)|}{T}_{m,\omega}\sqrt{\frac{L_{N}}{\ell_{\text{eff}}}}.
 \end{align*}
 Using this simplification we can calculate the integral over $x$ in Eq.~\eqref{eq:b_integral}
 \begin{align*}
  {b}_{\omega}  &= \frac{1}{2\pi c}\int_{d}^{\infty}dx\int_{0}^{\infty}d\omega'\sqrt{\frac{\omega'}{\omega}}\Bigg(\Big[  e^{-i\frac{\omega'}{c}d}e^{i\frac{\omega'-\omega}{c}\left(x-d\right)}+e^{-i\frac{\omega'}{c}d}e^{-i\frac{\omega'-\omega}{c}\left(x-d\right)}\Big]a_{\omega'}\\ 
  &+ \, 2ie^{i\frac{\omega'}{c}\ell_{c}}\frac{\cos\left(\frac{\omega'}{c}\ell_{\text{eff}}+\varphi_{r}(\omega')\right)+|r_{m}(\omega')|\cos\left(\frac{\omega'}{c}\ell_{\text{eff}}\right)}{|t_{m}(\omega')|}{T}_{m,\omega'}\sqrt{\frac{L_{N}}{\ell_{\text{eff}}}}e^{i\frac{\omega'-\omega}{c}\left(x-d\right)} \\ 
  &-\, \Big[- 2ie^{-i\frac{\omega'}{c}\ell_{c}}\frac{\cos\left(\frac{\omega'}{c}\ell_{\text{eff}}+\varphi_{r}(\omega')\right)+|r_{m}(\omega')|\cos\left(\frac{\omega'}{c}\ell_{\text{eff}}\right)}{|t_{m}(\omega)|}{T}_{m,\omega'}^{\ast}\sqrt{\frac{L_{N}}{\ell_{\text{eff}}}}e^{-i\frac{\omega'+\omega}{c}\left(x-d\right)}\\ 
  &+ \, e^{i\frac{\omega'}{c}d}e^{-i\frac{\omega'+\omega}{c}\left(x-d\right)}+e^{i\frac{\omega'}{c}d}e^{i\frac{\omega'+\omega}{c}\left(x-d\right)}\Big]a_{\omega'}^{\dagger}\Bigg)
 \end{align*}
 Knowing that:
 \begin{align*}
 \int_{0}^{\infty} dy~e^{\pm i\frac{\omega\pm \omega'}{c}y}=-\lim_{\epsilon \rightarrow 0}  \frac{c}{\left(\pm i(\omega\pm \omega')-\epsilon \right)}
 \end{align*}
and taking into account that we integrate only over positive frequencies, for the above integral we get 
\begin{eqnarray*}
{b}_{\omega}
&=&e^{-i\frac{\omega}{c}d}{a}_{\omega}
 +\frac{1}{\pi}\int_{0}^{\infty}d\omega'\;\sqrt{\frac{\omega'}{\omega}} \sqrt{\frac{L_{N}}{\ell_{\text{eff}}}}
  \frac{\cos\left(\frac{\omega'}{c}\ell_{\text{eff}}+\varphi_{r}(\omega')\right)+|r_{m}(\omega')|\cos\left(\frac{\omega'}{c}\ell_{\text{eff}}\right)}{|t_{m}(\omega')|}\\
&& \times \, \left(e^{i\frac{\omega'}{c}\ell_{c}}{T}_{m,\omega'}\lim_{\epsilon\rightarrow0}\frac{1}{\omega-\omega'-i\epsilon}{a}_{\omega'}
	+e^{-i\frac{\omega'}{c}\ell_{c}}{T}_{m,\omega'}^{\ast}\lim_{\epsilon\rightarrow0}\frac{1}{\omega+\omega'-i\epsilon}{a}_{\omega'}^{\dagger}\right).
\end{eqnarray*}

 \section{Commutation relation for the separated modes\label{app:commutation_satisfy}}
 
  In this appendix we verify the validity of the commutation relation~\eqref{eq:separated_com} using the separation~\eqref{eq:mode_relation_global_separate} with the coefficients~\eqref{eq:approximate_coef}:
  \begin{align}
  \nonumber
  \left[ a_{D}(\omega),a^{\dagger}_{D}(\omega')\right]&=\sum_{m=1}\sum_{m'=1}\alpha_{m_{\color{black}{-}}}(\omega)\alpha^{\ast}_{m'_{\color{black}{-}}}(\omega')\left[a_{m},a^{\dagger}_{m'}\right]\\ \nonumber
  &=\sum_{m=1}e^{-i(\omega-\omega')\frac{\ell_{c}}{c}}\frac{\textcolor{black}{\kappa}_{m}}{2\pi}\frac{1}{(\omega-\omega_{m})-i\frac{\textcolor{black}{\kappa}_{m}}{2}}\frac{1}{(\omega'-\omega_{m})+i\frac{\textcolor{black}{\kappa}_{m}}{2}}\\ \label{eq:app:a_D}
  &\approx\sum_{m=1}\frac{\textcolor{black}{\kappa}_{m}}{2\pi}\frac{1}{(\omega-\omega_{m})-i\frac{\textcolor{black}{\kappa}_{m}}{2}}\frac{1}{(\omega'-\omega_{m})+i\frac{\textcolor{black}{\kappa}_{m}}{2}}
  \end{align}
  For the corresponding continuous part we get:
  \begin{align*}
   \left[ a_{C}(\omega),a^{\dagger}_{C}(\omega'')\right]   =& \int_0^{\infty}  \int_0^{\infty}d\omega' d\tilde{\omega} \,\beta_{\color{black}{-}}(\omega, \omega') \beta^{\ast}_{\color{black}{-}}(\omega'', \tilde{\omega})\left[{b}_{\omega'},{b}^{\dagger}_{\tilde{\omega}}\right]\\
   =&\; \delta(\omega''-\omega)+ e^{i\frac{\omega}{c}d} \frac{1}{\pi} \left(e^{i\frac{\omega''}{c}\ell_{c}}{T}_{m,\omega''}\sqrt{\frac{L_{N}}{\ell_{\rm{eff}}}}(-1)^{m}\frac{-1+|r_{m}(\omega_{m})|}{|t_{m}(\omega_{m})|}\lim_{\epsilon\rightarrow 0}\frac{1}{\omega-\omega''-i\epsilon}\right)\\ 
   &+ e^{-i\frac{\omega''}{c}d}\frac{1}{\pi} \left(e^{-i\frac{\omega}{c}\ell_{c}}{T}^{\ast}_{m,\omega}\sqrt{\frac{L_{N}}{\ell_{\rm{eff}}}}(-1)^{m}\frac{-1+|r_{m}(\omega_{m})|}{|t_{m}(\omega_{m})|}\lim_{\epsilon\rightarrow 0}\frac{1}{\omega''-\omega+i\epsilon}\right)\\ 
   &+ \int^{\infty}_{0}d\omega' \frac{1}{\pi^{2}}e^{i(\omega''-\omega)\frac{\ell_{c}}{c}}\frac{L_{N}}{\ell_{\rm{eff}}}\left( \frac{-1+|r_{m}(\omega_{m})|}{|t_{m}(\omega_{m})|}\right)^{2}{T}^{\ast}_{m,\omega}{T}_{m,\omega''}\\
  & \times \lim_{\epsilon\rightarrow 0}\frac{1}{\omega'-\omega+i\epsilon}\lim_{\epsilon\rightarrow 0}\frac{1}{\omega'-\omega''-i\epsilon}.
  \end{align*}
  We proceed by applying the following approximation, which is justified by the fact that we consider a cavity of high reflectivity:
  \begin{align}
  \label{app:eq:good_cavity_approx}
\frac{-1+|r_{m}(\omega_{m})|}{|t_{m}(\omega_{m})|}=\frac{-1+e^{-\textcolor{black}{\kappa}_{m}\frac{\ell_{\text{eff}}}{c}}}{\sqrt{1-e^{-2\textcolor{black}{\kappa}_{m}\frac{\ell_{\text{eff}}}{c}}}}= \frac{-1+1-\textcolor{black}{\kappa}_{m}\frac{\ell_{\text{eff}}}{c}+\mathcal{O}\left(\left(\textcolor{black}{\kappa}_{m}\frac{\ell_{\text{eff}}}{c}\right)^{2}\right)}{\sqrt{1-1+2\textcolor{black}{\kappa}_{m}\frac{\ell_{\text{eff}}}{c}-\mathcal{O}\left(\left(\textcolor{black}{\kappa}_{m}\frac{\ell_{\text{eff}}}{c}\right)^{2}\right)}}\approx-\frac{\textcolor{black}{\kappa}_{m}\frac{\ell_{\text{eff}}}{c}}{\sqrt{2\textcolor{black}{\kappa}_{m}\frac{\ell_{\text{eff}}}{c}}}=-\sqrt{\frac{\textcolor{black}{\kappa}_{m}}{2}\frac{\ell_{\text{eff}}}{c}},
 \end{align}
 where the terms of the order of $\left(\textcolor{black}{\kappa}_{m}\frac{\ell_{\rm{eff}}}{c}\right)^{2}$ and higher are neglected based on the above assumption. With this assumption and using the explicit expression of $\mathcal{T}_{m}(\omega)$(since $T_{m,\omega}\approx \mathcal{T}_{m}(\omega)$) the commutator becomes:
 \begin{align}
 \nonumber
  [ a_{C}(\omega),a^{\dagger}_{C}(\omega'')] \approx& \, \delta(\omega''-\omega)- \hspace{-0.1cm} \sum_{m}(-1)^{m}\frac{\textcolor{black}{\kappa}_{m}}{2\pi}\lim_{\epsilon\rightarrow 0}\frac{1}{\omega-\omega''-i\epsilon}\hspace{-0.1cm}\\ \nonumber
&  \times \left(\hspace{-0.1cm}e^{i\frac{\omega}{c}d} e^{i\frac{\omega''}{c}\ell_{c}}\frac{1}{\omega''-\omega_{m}+i\frac{\textcolor{black}{\kappa}_{m}}{2}}-e^{-i\frac{\omega''}{c}d}e^{-i\frac{\omega}{c}\ell_{c}}\frac{1}{\omega-\omega_{m}-i\frac{\textcolor{black}{\kappa}_{m}}{2}}\hspace{-0.1cm}\right) \\ \nonumber
 &  + \frac{1}{\pi^{2}}e^{i(\omega''-\omega)\frac{\ell_{c}}{c}}\sum_{m}\frac{\textcolor{black}{\kappa}_{m}^{2}}{4}\frac{1}{\omega-\omega_{m}-i\frac{\textcolor{black}{\kappa}_{m}}{2}}\frac{1}{\omega''-\omega_{m}+i\frac{\textcolor{black}{\kappa}_{m}}{2}}\\ \nonumber
 &\times \int^{\infty}_{0}d\omega' \lim_{\epsilon\rightarrow 0}\frac{1}{\omega'-\omega+i\epsilon}\lim_{\epsilon\rightarrow 0}\frac{1}{\omega'-\omega''-i\epsilon}\\ \nonumber
   =&\, \delta(\omega''-\omega)- \sum_{m}\frac{\textcolor{black}{\kappa}_{m}}{2\pi}\lim_{\epsilon\rightarrow 0}\frac{1}{\omega-\omega''-i\epsilon}\left(\frac{1}{\omega''-\omega_{m}+i\frac{\textcolor{black}{\kappa}_{m}}{2}}-\frac{1}{\omega-\omega_{m}-i\frac{\textcolor{black}{\kappa}_{m}}{2}}\right)\\ \nonumber
   & + \frac{1}{\pi^{2}}e^{i(\omega''-\omega)\frac{\ell_{c}}{c}}\sum_{m}\frac{\textcolor{black}{\kappa}_{m}^{2}}{4}\frac{1}{\omega-\omega_{m}-i\frac{\textcolor{black}{\kappa}_{m}}{2}}\frac{1}{\omega''-\omega_{m}+i\frac{\textcolor{black}{\kappa}_{m}}{2}}\\ \nonumber
   &\times \int^{\infty}_{0}d\omega' \lim_{\epsilon\rightarrow 0}\frac{1}{\omega'-\omega+i\epsilon}\lim_{\epsilon\rightarrow 0}\frac{1}{\omega'-\omega''-i\epsilon}\\  \label{eq:app:a_c}
  & \approx \delta(\omega''-\omega)-  \sum_{m}\frac{\textcolor{black}{\kappa}_{m}}{2\pi}\frac{1}{\left(\omega''-\omega_{m}+i\frac{\textcolor{black}{\kappa}_{m}}{2}\right)\left(\omega-\omega_{m}-i\frac{\textcolor{black}{\kappa}_{m}}{2}\right)},
\end{align} 
 where we have neglected the terms of the order of $\left(\textcolor{black}{\kappa}_{m}/\omega_{m}\right)^{2}$ since we consider cavities of high quality, i.e., $\textcolor{black}{\kappa}_{m}/\omega_{m}\ll 1$. We also evaluated the following term at the resonance frequency $\omega_{m}$: $e^{i\frac{\omega}{c}d}e^{i\frac{\omega''}{c}\ell_{c}}=e^{i\frac{\omega_{m}}{c}d}e^{i\frac{\omega_{m}}{c}\ell_{c}}=e^{i\frac{\omega_{m}}{c}(\ell_{\rm{eff}}-\ell_{c})}e^{i\frac{\omega_{m}}{c}\ell_{c}}=(-1)^{m}$.
 
 If we sum equations~\eqref{eq:app:a_D} and~\eqref{eq:app:a_c} we get the required commutation relation:
 \begin{eqnarray}
 \left[ a_{\omega},a^{\dagger}_{\omega'}\right]&=&\left[ a_D(\omega),a^{\dagger}_D(\omega')\right]+\left[ a_C(\omega),a^{\dagger}_C(\omega')\right]=\delta(\omega'-\omega).
 \end{eqnarray}

 \section{Hamiltonian representation \label{app:Hamiltonian}}

In this appendix we derive the Hamiltonian for the inside-outside representation, starting from Hamiltonian~\eqref{eq:global_Hamiltonian} derived for the true modes. Using the relation~\eqref{eq:mode_relation_global_separate} we can write the following:
  \begin{align*}
  H=&\int_{0}^{\infty}d\omega~\hbar \omega  {a}^{\dagger}_{\omega}{a}_{\omega}\\
  =&\int^{\infty}_{0}d\omega \hbar \omega \Bigg[\sum_{m=1}\sum_{m'=1}\alpha_{m_{\color{black}{-}}}(\omega)\alpha^{\ast}_{m'_{\color{black}{-}}}(\omega){a}^{\dagger}_{m'}{a}_{m}+\sum_{m=1}\int^{\infty}_{0} d\omega' \Big(\alpha^{\ast}_{m_{\color{black}{-}}}(\omega)\beta_{{\color{black}{-}}}(\omega,\omega'){a}^{\dagger}_{m}{b}_{\omega'}\\
  &+\alpha_{m_{\color{black}{-}}}(\omega)\beta^{\ast}_{{\color{black}{-}}}(\omega,\omega'){b}^{\dagger}_{\omega'}{a}_{m}\Big)+\int_{0}^{\infty} d\omega' d\omega'' \beta^{\ast}_{{\color{black}{-}}}(\omega,\omega')\beta_{{\color{black}{-}}}(\omega,\omega'') {b}^{\dagger}_{\omega'}{b}_{\omega''}\Bigg].
  \end{align*}
 We proceed by analyzing each term of the above equation separately:
 \begin{align}
 \label{eq:cavity_part}
 H_{C} = \int^{\infty}_{0}d\omega \hbar \omega \sum_{m=1}\sum_{m'=1}\alpha_{m_{\color{black}{-}}}(\omega)\alpha^{\ast}_{m'_{\color{black}{-}}}(\omega){a}^{\dagger}_{m}{a}_{m'}
 \end{align}
 Since we are considering a regime where individual resonance peaks of the cavity are well separated, we can apply the following estimation: $\alpha_{m_{\color{black}{-}}}(\omega)\alpha^{\ast}_{m'_{\color{black}{-}}}(\omega)\approx \delta_{mm'}\lvert \alpha_{m_{\color{black}{-}}}(\omega)\rvert^{2}$, which follows from the assumption $\mathcal{T}_{m}(\omega)\mathcal{T}^{\ast}_{m'}(\omega)=\delta_{mm'}$. Taking this into account the expression~\eqref{eq:cavity_part} becomes
 \begin{align}
H_{C} =  \int^{\infty}_{0}d\omega \hbar \omega \sum_{m=1}\lvert \alpha_{m_{\color{black}{-}}}(\omega)\rvert^{2}{a}^{\dagger}_{m}{a}_{m},
 \end{align}
 where the following integral can be evaluated by using the corresponding complex integral, assuming that $\textcolor{black}{\kappa}_{m}/\omega_{m} \ll 1$: 
\begin{align}
\label{app:eq:real_int}
 \int^{\infty}_{0}d\omega \, \omega \lvert \alpha_{m_{\color{black}{-}}}(\omega)\rvert^{2}=\int^{\infty}_{0}d\omega \, \frac{\textcolor{black}{\kappa}_{m}}{2 \pi}\frac{\omega}{(\omega-\omega_{m})^{2}+\left(\frac{\textcolor{black}{\kappa}_{m}}{2}\right)^{2}},
\end{align}
the corresponding complex integral being $\displaystyle \oint_{C} dz~\dfrac{z+b}{z^{2}+a}$, integrated in the upper half of the complex plane. Taking this into account the integral in~\eqref{app:eq:real_int} becomes 
 \begin{align*}
 \int^{\infty}_{0}d\omega \, \omega \lvert \alpha_{n_{\color{black}{-}}}(\omega)\rvert^{2}\approx \omega_{m},
 \end{align*}
 which leads to 
  \begin{align}
 H_{C} =\sum_{m=1}\hbar \omega_{m}{a}^{\dagger}_{m}{a}_{m}.
 \end{align}
 
 Next we evaluate the continuous part of the Hamiltonian describing the reservoir:
 \begin{align}
 \label{eq:Hamiltonian_res_part}
 H_{R} = &\int_{0}^{\infty}d\omega\int_{0}^{\infty}d\omega'\int_{0}^{\infty}d\omega''\;\hbar\omega \,\beta_{{\color{black}{-}}}^{\ast}(\omega,\omega')\beta_{{\color{black}{-}}}(\omega,\omega'') {b}^{\dagger}_{\omega'}{b}_{\omega''}
 \end{align}
  Using the approximation in~\eqref{app:eq:good_cavity_approx} the expression of $\beta_{1}(\omega,\omega')$ can be approximated as follows:
  \begin{equation}
  \label{app:eq:beta_app}
 \begin{split}
 \beta_{{\color{black}{-}}}(\omega,\omega')&\approx e^{i\frac{\omega'}{c}d}\delta(\omega-\omega')-\frac{1}{\pi}e^{-i\frac{\omega}{c}\ell_{c}}\mathcal{T}^{\ast}_m(\omega)(-1)^{m}\sqrt{\frac{\textcolor{black}{\kappa}_{m}\ell_{\rm{eff}}}{2c}}\sqrt{\frac{L_{N}}{\ell_{\rm{eff}}}}\lim_{\epsilon \rightarrow 0} \frac{1}{\omega'-\omega+i\epsilon}\\
 &=e^{i\frac{\omega'}{c}d}\delta(\omega-\omega')-\frac{1}{\sqrt{\pi}}e^{-i\frac{\omega}{c}\ell_{c}}\sum_{m}\sqrt{\frac{\textcolor{black}{\kappa}_{m}}{2\pi}}\frac{1}{\omega-\omega_{m}-i\frac{\textcolor{black}{\kappa}_{m}}{2}}(-1)^{m}\sqrt{\frac{\textcolor{black}{\kappa}_{m}}{2}}\lim_{\epsilon \rightarrow 0} \frac{1}{\omega'-\omega+i\epsilon}\\
 &=e^{i\frac{\omega'}{c}d}\delta(\omega-\omega')-\frac{1}{\sqrt{\pi}}\sum_{m}\alpha_{m_1}(\omega)(-1)^{m}\sqrt{\frac{\textcolor{black}{\kappa}_{m}}{2}}\lim_{\epsilon \rightarrow 0} \frac{1}{\omega'-\omega+i\epsilon}
 \end{split}
 \end{equation}
Taking this into account, the Hamiltonian in~\eqref{eq:Hamiltonian_res_part} becomes:
 \begin{align*}
H_{R} \approx &\int_{0}^{\infty}d\omega\:\hbar\omega{b}_{\omega}^{\dagger}{b}_{\omega}+\frac{1}{2\pi}\int_{0}^{\infty}d\omega' \int_{0}^{\infty} d\omega '' \;\hbar \Bigg(\lim_{\epsilon\rightarrow0}\frac{\textcolor{black}{\kappa}_{m}}{\omega''-\omega'+i\epsilon} \frac{-\omega''\left(\omega_{m}+i\frac{\textcolor{black}{\kappa}_{m}}{2}\right)+\omega'\left( \omega_{m}-i\frac{\textcolor{black}{\kappa}_{m}}{2}\right)}{\left(\omega'-\omega_{m}-i\frac{\textcolor{black}{\kappa}_{m}}{2}\right)\left(\omega''-\omega_{m}+i\frac{\textcolor{black}{\kappa}_{m}}{2}\right)}\\
 &+\int_{0}^{\infty}d\omega\;\omega\frac{\textcolor{black}{\kappa}^{2}_{m}}{(\omega-\omega_{m})^{2}+\left(\frac{\textcolor{black}{\kappa}_{m}}{2}\right)^{2}}\lim_{\epsilon\rightarrow0}\frac{1}{\omega'-\omega-i\epsilon}\lim_{\epsilon\rightarrow0}\frac{1}{\omega''-\omega+i\epsilon}\Bigg){b}^{\dagger}_{\omega'}{b}_{\omega''}.
 \end{align*}
 By estimating the integral in the last term via corresponding complex plane integral (similar to eq.~\eqref{app:eq:real_int}), and neglecting the terms of the order $(\textcolor{black}{\kappa}_{m}/{\omega_{m}})^{2}$, the above expression reduces to the following:
   \begin{align*}
H_{R}= \int_{0}^{\infty}d\omega\:\hbar\omega {b}_{\omega}^{\dagger}{b}_{\omega}. 
 \end{align*}
 
 Finally, we evaluate the term describing the interaction between perfect cavity and reservoir modes: 
 \begin{align}
 \label{app:eq:coupling_Hamiltonian}
H_{CR} &=   \int_{0}^{\infty}d\omega \,\hbar \omega \sum_{m=1}\int^{\infty}_{0} d\omega' \left(\alpha^{\ast}_{m_{\color{black}{-}}}(\omega)\beta_{{\color{black}{-}}}(\omega,\omega'){a}^{\dagger}_{m}{b}_{\omega'}+\alpha_{m_{\color{black}{-}}}(\omega)\beta^{\ast}_{{\color{black}{-}}}(\omega,\omega'){b}^{\dagger}_{\omega'}{a}_{m}\right).
 \end{align}
 To simplify this expression we  again use the approximated expression~\eqref{app:eq:beta_app} for $\beta_{{\color{black}{-}}}(\omega,\omega')$, leading to an integral of the form
 \begin{align*}
\mathcal{I}_{1}&= \int_{0}^{\infty}d\omega \,\hbar \omega \sum_{m=1}\int^{\infty}_{0} d\omega' \alpha^{\ast}_{m_{\color{black}{-}}}(\omega)\beta_{{\color{black}{-}}}(\omega,\omega'){a}^{\dagger}_{m}{b}_{\omega'}=\int_{0}^{\infty}d\omega  \sum_{m=1} \hbar \omega \, \alpha^{\ast}_{m_{\color{black}{-}}}(\omega)e^{i\frac{\omega}{c}d}a^{\dagger}_{m}{b}_{\omega}\\
&- \int_{0}^{\infty}d\omega  \sum_{m=1}\int^{\infty}_{0} d\omega' \sum_{m'=1} \, \hbar \omega \, (-1)^{m'}  \alpha^{\ast}_{m_{\color{black}{-}}}(\omega)\alpha_{m'_{\color{black}{-}}}(\omega) \sqrt{\frac{\textcolor{black}{\kappa}_{m'}}{2\pi}}\lim_{\epsilon\rightarrow0}\frac{1}{\omega'-\omega+i\epsilon} a^{\dagger}_{m}{b}_{\omega'}\\
&=\int_{0}^{\infty}d\omega  \sum_{m=1} \hbar \omega \, \alpha^{\ast}_{m_{\color{black}{-}}}(\omega)e^{i\frac{\omega}{c}d}a^{\dagger}_{m}{b}_{\omega}\\
&- \int_{0}^{\infty}d\omega  \sum_{m=1}\int^{\infty}_{0} d\omega'  \, \hbar \omega \, (-1)^{m}  \lvert \alpha_{m_{\color{black}{-}}}(\omega)\rvert^{2} \sqrt{\frac{\textcolor{black}{\kappa}_{m}}{2\pi}}\lim_{\epsilon\rightarrow0}\frac{1}{\omega'-\omega+i\epsilon} a^{\dagger}_{m}{b}_{\omega'}\\
 \end{align*}
 The last term can be evaluated similar to integral in~\eqref{app:eq:real_int}, leading to
  \begin{align*}
 \mathcal{I}_{1}&=\int_{0}^{\infty}d\omega  \sum_{m=1} \hbar \omega \, \alpha^{\ast}_{m_{\color{black}{-}}}(\omega)e^{i\frac{\omega}{c}d}a^{\dagger}_{m}{b}_{\omega}- \int_{0}^{\infty}d\omega  \sum_{m=1}  \, \hbar   (-1)^{m}   \sqrt{\frac{\textcolor{black}{\kappa}_{m}}{2\pi}}\frac{\omega_{m}-i\frac{\textcolor{black}{\kappa}_{m}}{2}}{\omega-\omega_{m}+i\frac{\textcolor{black}{\kappa}_{m}}{2}} a^{\dagger}_{m}{b}_{\omega}\\
 &=\int_{0}^{\infty}d\omega  \sum_{m=1} \hbar \Bigg( \omega \, e^{i\frac{\omega}{c}\ell_{\rm{eff}}}\sqrt{\frac{\textcolor{black}{\kappa}_{m}}{2\pi}}\frac{1}{\omega-\omega_{m}+i\frac{\textcolor{black}{\kappa}_{m}}{2}} -  (-1)^{m}   \sqrt{\frac{\textcolor{black}{\kappa}_{m}}{2\pi}}\frac{\omega_{m}-i\frac{\textcolor{black}{\kappa}_{m}}{2}}{\omega-\omega_{m}+i\frac{\textcolor{black}{\kappa}_{m}}{2}}\Bigg) a^{\dagger}_{m}{b}_{\omega}\\
 &\approx \int_{0}^{\infty}d\omega  \sum_{m=1} \hbar \Bigg(  (-1)^{m}\sqrt{\frac{\textcolor{black}{\kappa}_{m}}{2\pi}}\frac{\omega }{\omega-\omega_{m}+i\frac{\textcolor{black}{\kappa}_{m}}{2}} - (-1)^{m}   \sqrt{\frac{\textcolor{black}{\kappa}_{m}}{2\pi}}\frac{\omega_{m}-i\frac{\textcolor{black}{\kappa}_{m}}{2}}{\omega-\omega_{m}+i\frac{\textcolor{black}{\kappa}_{m}}{2}}\Bigg) a^{\dagger}_{m}{b}_{\omega}\\
 &=\int_{0}^{\infty}d\omega  \sum_{m=1} \hbar e^{i\frac{\omega}{c}\ell_{\rm{eff}}}\sqrt{\frac{\textcolor{black}{\kappa}_{m}}{2\pi}}\text{sinc}\left[(\omega-\omega_{m})\frac{\ell_{\rm{eff}}}{c}\right] a^{\dagger}_{m}{b}_{\omega},
 \end{align*}
 where when writing the last term we recovered the terms we had previously evaluated at the resonance frequency $\omega_{m}$, i.e., $\text{sinc}\left[(\omega-\omega_{m})\frac{\ell_{\rm{eff}}}{c}\right]$ and $e^{i\frac{\omega}{c}\ell_{\rm{eff}}}$.
  Further, by defining a coupling function as follows:
  \begin{align}
  \textcolor{black}{V}_{m}(\omega)&=-ie^{-i\frac{\omega}{c}\ell_{\rm{eff}}}\sqrt{\frac{\textcolor{black}{\kappa}_{m}}{2\pi}}\text{sinc}\left[(\omega-\omega_{m})\frac{\ell_{\rm{eff}}}{c}\right],
  \end{align}
the full Hamiltonian in terms of separated inside-outside modes becomes:
\begin{align*}
H = \sum_{m=1}\hbar \omega_{m}{a}^{\dagger}_{m}{a}_{m}+\int_{0}^{\infty}d\omega\:\hbar\omega {b}_{\omega}^{\dagger}{b}_{\omega}+i\hbar \sum_{m=1}\int^{\infty}_{0} d\omega \, \left(\textcolor{black}{V}_{m}(\omega){b}^{\dagger}_{\omega}{a}_{m}- \textcolor{black}{V}^{\ast}(\omega){a}^{\dagger}_{m}{b}_{\omega}\right).
\end{align*}

 \color{black}{
 \section{Single photon spatial distribution \label{app:photon_in_space}}
 
 In this appendix, we demonstrate that the definition in Eq.~\eqref{eq:number_space} can be interpreted as a photon number operator. As discussed in Ref.~\cite{garrison2008quantum}, the attempts to define a local photon number operator in a finite volume $V$ showed that, in general, such an operator can not be defined since it does not satisfy the commutation relation for nonoverlapping volumes. However, in the particular system discussed here, where linearly polarized light produced from a high-finesse cavity propagates in one-dimensional space, we are able to define such an operator. In order to show that the definition in Eq.~\eqref{eq:number_space} is an observable, we evaluate the following commutator 
 \begin{align}
 \label{eq:commutator_x}
\left[b_{x}^{\dagger}b_{x},b_{x'}^{\dagger}b_{x'}\right]=&\,  \frac{1}{(c\textcolor{black}{\kappa}_{m})^{2}}\int d\omega d\omega'd\omega''d\tilde{\omega}\textcolor{black}{V}_{m}(\omega)\textcolor{black}{V}_{m}^{\ast}(\omega')\textcolor{black}{V}_{m}(\omega'')\textcolor{black}{V}_{m}^{\ast}(\tilde{\omega})\\ \nonumber
& \times e^{-i\frac{\omega}{c}x}e^{i\frac{\omega'}{c}x}e^{-i\frac{\omega''}{c}x'}e^{i\frac{\tilde{\omega}}{c}x'}\left[b_{\omega}^{\dagger}b_{\omega'},b_{\omega''}^{\dagger}b_{\tilde{\omega}}\right]. 
\end{align}
The commutator appearing on the right side of the equation~\eqref{eq:commutator_x} can be simplified as follows 
\begin{align*}
\left[b_{\omega}^{\dagger}b_{\omega'},b_{\omega''}^{\dagger}b_{\tilde{\omega}}\right] &= b_{\omega}^{\dagger}b_{\omega'}b_{\omega''}^{\dagger}b_{\tilde{\omega}}-b_{\omega''}^{\dagger}b_{\tilde{\omega}} b_{\omega}^{\dagger}b_{\omega'} = b_{\omega}^{\dagger}\left(\delta(\omega'-\omega'')+b_{\omega''}^{\dagger}b_{\omega'}\right)b_{\tilde{\omega}}-b_{\omega''}^{\dagger}b_{\tilde{\omega}} b_{\omega}^{\dagger}b_{\omega'} \\
&=b_{\omega}^{\dagger}b_{\tilde{\omega}}\delta(\omega'-\omega'')+b_{\omega}^{\dagger}b_{\omega''}^{\dagger}b_{\omega'}b_{\tilde{\omega}}-b_{\omega''}^{\dagger}b_{\tilde{\omega}} b_{\omega}^{\dagger}b_{\omega'}=b_{\omega}^{\dagger}b_{\tilde{\omega}}\delta(\omega'-\omega'')-b_{\omega''}^{\dagger}b_{\omega'}\delta(\omega-\tilde{\omega}).
\end{align*}
Taking this into account Eq.~\eqref{eq:commutator_x} becomes
\begin{eqnarray*}
\left[b_{x}^{\dagger}b_{x},b_{x'}^{\dagger}b_{x'}\right]&=&\frac{1}{(c\textcolor{black}{\kappa}_{m})^{2}}\int d\omega d\omega'd\tilde{\omega}\textcolor{black}{V}_{m}(\omega)\textcolor{black}{V}_{m}^{\ast}(\tilde{\omega})|\textcolor{black}{V}_{m}(\omega')|^{2}e^{i\frac{\omega'}{c}(x-x')}e^{-i\frac{\omega}{c}x}e^{i\frac{\tilde{\omega}}{c}x'}b_{\omega}^{\dagger}b_{\tilde{\omega}}\\
&&-\frac{1}{(c\textcolor{black}{\kappa}_{m})^{2}}\int d\omega d\omega'd\omega''\textcolor{black}{V}_{m}^{\ast}(\omega')\textcolor{black}{V}_{m}(\omega'')|\textcolor{black}{V}_{m}({\omega})|^{2}e^{-i\frac{\omega}{c}(x-x')}e^{i\frac{\omega'}{c}x}e^{-i\frac{\omega''}{c}x'}b_{\omega''}^{\dagger}b_{\omega'}
\end{eqnarray*}
To evaluate these integrals, below we estimate the value of the integral 
\begin{align}
\nonumber
\mathcal{I} & = \int_{0}^{\infty}d\omega'|\textcolor{black}{V}_{m}(\omega')|^{2}e^{i\frac{\omega'}{c}(x-x')}=\frac{{\textcolor{black}{\kappa}_{m}}}{2\pi}\left(\frac{c}{\ell_{\text{eff}}}\right)^{2}\int_{0}^{\infty}d\omega'\frac{\sin^{2}\left(\omega'-\omega_{m}\right)\frac{\ell_{\text{eff}}}{c}}{(\omega'-\omega_{m})^{2}}e^{i\frac{\omega'}{c}(x-x')}\\\nonumber
&=\frac{{\textcolor{black}{\kappa}_{m}}}{2\pi}\left(\frac{c}{\ell_{\text{eff}}}\right)^{2}\frac{1}{2}\int_{0}^{\infty}d\omega'\frac{1-\cos2(\omega'-\omega_{m})\frac{\ell_{\text{eff}}}{c}}{(\omega'-\omega_{m})^{2}}e^{i\frac{\omega'}{c}(x-x')}\\ \label{eq:fullint}
&=\frac{{\textcolor{black}{\kappa}_{m}}}{2\pi}\left(\frac{c}{\ell_{\text{eff}}}\right)^{2}\frac{1}{2}e^{i\frac{\omega_{m}}{c}(x-x')}\left[\int_{0}^{\infty}\hspace{-0.25cm}d\omega'\frac{e^{i(\omega'-\omega_{m})\frac{(x-x')}{c}}}{(\omega'-\omega_{m})^{2}}\right.\\ \nonumber 
&-\left.\frac{1}{2}\left(\int_{0}^{\infty}\hspace{-0.25cm}d\omega'\frac{e^{i\frac{(\omega'-\omega_{m})}{c}(x-x'+2\ell_{\text{eff}})}}{(\omega'-\omega_{m})^{2}}+\int_{0}^{\infty}\hspace{-0.25cm}d\omega\frac{e^{i\frac{(\omega'-\omega_{m})}{c}(x-x'-2\ell_{\text{eff}})}}{(\omega'-\omega_{m})^{2}}\right)\right]
\end{align}
To calculate the individual terms of this integral, we can evaluate the corresponding integrals in complex plane, i.e., we evaluate integrals of the following form:
\begin{align*}
\int_{-\omega_{m}}^{\infty}dz\frac{e^{i\frac{z}{c}\zeta}}{z^{2}}\approx \int_{-\infty}^{\infty}dz\frac{e^{i\frac{z}{c}\zeta}}{z^{2}},
\end{align*}
where we have extended the limits of the integration, assuming that $c/\ell_{\text{eff}} \ll \omega_{m}$, which is satisfied when we consider high-finesse cavities, where the number of modes inside the cavity is high. Calculating the integrals in the complex plane one can show that 
\begin{align*}
\mathcal{I}=\frac{{\textcolor{black}{\kappa}_{m}}}{2\pi}\left(\frac{c}{\ell_{\text{eff}}}\right)^{2}\frac{1}{2}e^{i\frac{\omega_{m}}{c}(x-x')}\begin{cases}
\begin{array}{c}
0, \quad  x'<x-2\ell_{\text{eff}}\\
-\frac{\pi}{c}(x-x'-2\ell_{\text{eff}}),\quad x-2\ell_{\text{eff}}<x'<x \\
\frac{\pi}{c}(x-x'+2\ell_{\text{eff}}), \quad x<x'<x+2\ell_{\text{eff}} \\
0, \quad  x'>x+2\ell_{\text{eff}}
\end{array}
\end{cases}.
\end{align*}
Taking this result into account, we can conclude that commutator~\eqref{eq:commutator_x} is 0 when $|x'-x|>2\ell_{\text{eff}}$. Equivalently, if we formulate this condition in terms of the propagation time of the photon, we can write $x' = x+ct$, where $t>2\ell_{\text{eff}}/c$, which is known as the coarse-grained approximation, which assumes that the time of flight of a photon through the cavity is small compared with the time resolution of interest~\cite{vogel2006quantum}. Thus, at this limit, we can interpret  $b_{x}^{\dagger}b_{x}$ as the photon number density. 
}
\color{black}

\end{document}